\documentclass[lettersize,journal]{IEEEtran}
\usepackage{amsmath,amsfonts}
\usepackage{algorithmic}
\usepackage{algorithm}
\usepackage{array}
\usepackage[caption=false,font=normalsize,labelfont=sf,textfont=sf]{subfig}
\usepackage{textcomp}
\usepackage{stfloats}
\usepackage{url}
\usepackage{verbatim}
\usepackage{graphicx}
\usepackage{cite}
\usepackage{makecell}
\usepackage{tabularx}
\usepackage{adjustbox}
\usepackage{hyperref}
\usepackage{orcidlink}
\usepackage{wasysym}
\usepackage{booktabs}
\usepackage{xcolor}
\newcommand{\major}[1]{\textcolor{black}{#1}}
\newcommand{\minor}[1]{\textcolor{black}{#1}}
\UseRawInputEncoding

\newcommand{\commento}[1]{}

\begin{document}

\title{A Survey on Decentralized Identifiers and Verifiable Credentials}

\begin{comment}
    \author{Carlo Mazzocca \orcidlink{0000-0001-8949-2221},~\IEEEmembership{Member,~IEEE,} Abbas Acar \orcidlink{0000-0002-4891-160X}, Selcuk Uluagac \orcidlink{0000-0002-9823-3464},~\IEEEmembership{Senior Member,~IEEE}  Rebecca Montanari \orcidlink{0000-0002-3687-0361},~\IEEEmembership{Member,~IEEE,} Paolo Bellavista \orcidlink{0000-0003-0992-7948},~\IEEEmembership{Senior Member,~IEEE} \and Mauro Conti \orcidlink{0000-0002-3612-1934},~\IEEEmembership{Fellow,~IEEE}
\end{comment}
% \author{Carlo Mazzocca \orcidlink{0000-0001-8949-2221}, Abbas Acar \orcidlink{0000-0002-4891-160X}, Selcuk Uluagac \orcidlink{0000-0002-9823-3464},~\IEEEmembership{Senior Member,~IEEE,}  Rebecca Montanari \orcidlink{0000-0002-3687-0361}, Paolo Bellavista \orcidlink{0000-0003-0992-7948},~\IEEEmembership{Senior Member,~IEEE,} \and Mauro Conti \orcidlink{0000-0002-3612-1934},~\IEEEmembership{Fellow,~IEEE}

\author{Carlo Mazzocca \orcidlink{0000-0001-8949-2221}, Abbas Acar \orcidlink{0000-0002-4891-160X}, Selcuk Uluagac \orcidlink{0000-0002-9823-3464},  Rebecca Montanari \orcidlink{0000-0002-3687-0361}, Paolo Bellavista \orcidlink{0000-0003-0992-7948}, \and Mauro Conti \orcidlink{0000-0002-3612-1934}

        % <-this % stops a space
\thanks{Manuscript received X, X; revised X, X.}% <-this % stops a space
\thanks{
Carlo Mazzocca is with the Department of Information and Electrical Engineering and Applied Mathematics, University of Salerno, 84084
Fisciano, Italy (e-mail: cmazzocca@unisa.it).\\
Rebecca Montanari and Paolo Bellavista are with the Department of Computer Science and Engineering, University of Bologna, 40136 Bologna, Italy (e-mail: {name.surname}@unibo.it).\\
Abbas Acar and Selcuk Uluagac are with the Cyber-Physical Systems Security Lab, School of Computing and Information Science, Florida International University, 33174 Miami, Florida, United States (e-mail: {aacar, suluagac}@fiu.edu).\\
Mauro Conti is with the Department of Mathematics, University of Padua, 35131 Padua, Italy (e-mail: conti@math.unipd.it).}}

\markboth{This article has been accepted for publication in IEEE Communications Surveys \& Tutorials: 10.1109/COMST.2025.3543197}%
{Shell \MakeLowercase{\textit{et al.}}: A Sample Article Using IEEEtran.cls for IEEE Journals}


\maketitle

\begin{abstract}
Digital identity has always been \minor{one of} the keystones for implementing secure and trustworthy communications among parties. The ever-evolving digital landscape has undergone \minor{numerous} technological transformations that have profoundly \minor{reshaped} digital identity management, \minor{leading to a major} shift from centralized to decentralized identity models.
The \minor{latest stage} of this \minor{evolution} is represented by the emerging \major{paradigm of} Self-Sovereign Identity (SSI), which gives \minor{identity owners} full control over their data. SSI leverages Decentralized Identifiers (DIDs) and Verifiable Credentials (VCs), which have been recently standardized by the World Wide Web Consortium (W3C). These technologies have the potential to build more secure and decentralized digital identity systems, significantly strengthening communication security \minor{in scenarios involving} many distributed participants. It is worth noting that \minor{use} of DIDs and VCs \major{is not limited to individuals but} \minor{extends to} a \major{wide} range of entities including cloud, edge, and Internet of Things (IoT) resources. 
However, due to their novelty, existing literature lacks a comprehensive survey on \major{DIDs and VCs beyond the scope of SSI}. This paper \major{fills \minor{this} gap by providing} a comprehensive overview of \major{DIDs and VCs} from \major{multiple} perspectives. 
\major{It identifies key security threats and mitigation strategies, \minor{analyzes} available implementations to guide practitioners in making informed decisions, \minor{and reviews the adoption of these technologies across various application domains}.  Moreover, \major{it} \minor{also examines} related regulations, projects, and consortiums emerging worldwide.} Finally, \major{it discusses the} \minor{primary} challenges hindering their real-world adoption and \minor{outlines} future research directions. 

\end{abstract}


\section{Introduction}

The recent proliferation of digital services \minor{accessible} through a network connection has \minor{led to} an unprecedented increase in the number of digital identities \cite{grassi2017digital}. As a result, the vast majority of the world's population owns at least one digital identity, which serves as the key to unlocking a multitude of online capabilities and services. However, the concept of digital identity extends far beyond the identification of human beings \cite{10.1007/978-3-319-50011-9_2}. With the widespread adoption of the Internet of Things (IoT) and the power of the 5th generation (5G) networks, coupled with the upcoming advent of the 6th generation (6G) \minor{networks}, there has been a remarkable increase in the number of connected devices \cite{8972389}. These devices require unique digital identities \minor{to} enable their participation in the digital ecosystem, \minor{such as} establishing secure communications. 

\minor{Digital identification} has always been a primary concern, as \minor{evidenced} by the \minor{numerous} solutions that have \minor{emerged} over the years. Indeed, the way an entity proves \minor{ownership} of a digital identity can be affected by a wide range of vulnerabilities \cite{OLIVERO2020102492}. 
Centralized identity providers represent \minor{a major weakness} of traditional identification methods.
Information stored in centralized data repositories \minor{can} potentially lead to serious data breaches, \minor{resulting in significant} loss of personal data and \minor{damage to stakeholders' reputations}  \cite{ALNEYADI2016137, shabtai2012survey}. 
In addition, centralized identity management systems \minor{rely} on specific identity service nodes, \minor{making them susceptible to the} single point of failure problem \cite{8869262}. 

The advent of federated identity systems has addressed some of these weaknesses along with scalability needs \cite{9869618}. The principal advantage of this identity model lies in empowering users to seamlessly access multiple services using the same identity. This identification paradigm relies on mutual trust among various parties, wherein verification is distributed across all identification systems or \minor{entities} \major{that mutually accept the standards employed by each system}. In a typical federated scenario, users authenticate themselves against a singular authority, referred to as the identity provider, 
\major{allowing} them to access all other applications on its behalf. Despite representing an advance,
federated identity still raises significant concerns regarding user privacy. Specifically, this model hinges on a certain level of trust in identity providers, and individuals retain limited control over their data.

Therefore, the increasing use of online services, the growing decentralization, as well as the \minor{rising} awareness of the drawbacks of traditional methods are paving the way to more secure and privacy-preserving methods. In this direction,
\minor{supported} by the current laws and regulations such as the European General Data Protection Regulation (GDPR) \cite{GDPR}, the concept of \textit{Self-Sovereign Identity} (SSI) \cite{MUHLE201880} is gaining \minor{significant} interest from both the academic and industrial worlds. 

SSI is based on the idea that an individual \minor{should} have full control \minor{over} their information without being forced to outsource data to any centralized authority or third party. With an SSI, users can \minor{directly} store their identity data and determine how much of \minor{them} they wish to share. \minor{In this way}, they \minor{can} decide with whom they share their data. A fundamental achievement of SSI is the \minor{ability for users} to present their trusted credentials to a third party without \minor{needing} an intermediary. This process is enabled through the ownership and control of Decentralized Identifiers (DIDs) \cite{did}, \minor{which} define a globally unique and cryptographic identifier scheme. Each DID is associated with a DID Document, \minor{containing} publicly available information (e.g., public keys), stored on shared \minor{and verifiable} data sources such as Distributed Ledger Technologies (DLTs). By presenting a DID, an individual can obtain Verifiable Credentials (VCs) \cite{vcs}. These VCs \minor{include} claims regarding the DID \minor{that} can be verified by an external party without \minor{requiring} direct engagement with the VC issuer. For example, an individual applying for a job online could \minor{present} a digitally signed credential from their university, attesting to their acquisition of a bachelor's degree and residency in the relevant country in which they are applying.

\begin{table*}[t!]
  \centering
  \caption{\major{Summary} of Surveys in the Field.}
  \label{tab:comparison1}
  
  \begin{adjustbox}{width=\textwidth}
    \begin{tabular}{>{\centering\arraybackslash}m{2cm}>{\centering\arraybackslash}m{2.5cm} m{0.67\linewidth}c} % Adjust the column width to fit the table
      \hline
      \textbf{Year} & \textbf{Ref.} & \textbf{Summary of the Work} \\
      \hline
      2021 & \major{Cucko et al.} \cite{9558805} & A systematic mapping study on SSI. The \major{study} outlines that DIDs and VCs \minor{are} discussed in most papers addressing SSI. The paper mainly provides insights into trends and demographics of SSI works. \\
      \hline
      2021 & \major{Soltani et al.} \cite{Soltani2021} & A survey on \major{the} SSI ecosystem. The work \major{mainly reviews literature related to SSI. It reports a few available implementations and briefly discusses related regulations in Europe and the United States.} \\
      \hline
      2022 & \major{Bai et al.} \cite{9881610} & A concise survey on the use of blockchain in SSI. The concept of DIDs and VCs is only presented as the foundational technologies of SSI, without delving into an examination of relevant research papers. \\
      \hline
      2022 & \major{Schardong et al.} \cite{s22155641} & A systematic review and mapping of theoretical and practical advances in SSI. The paper comprehensively discusses how DIDs and VCs can be employed in SSI systems, \major{and also considers a few other application domains}. \\
      \hline
        2023 & \major{Ernstberger et al.} \cite{Ernstberger} & \major{A systematization of knowledge on data sovereignty, with a focus on decentralized identity, decentralized access control, and policy-compliant decentralized computation. The paper also outlines key security and privacy properties, along with open challenges in the field.} \\
      \hline
        2024 & \major{Krul et al.} \cite{krul2024sok} & \major{A systemization of knowledge based on trust requirements and assumptions of SSI elements. The work identifies threats in SSI, potential mitigation, available implementations, and some challenges.} \\
      \hline
        2024 & \major{Tan et al.} \cite{tan} & \major{A systematic review and analysis of SSI solutions based on a set of research questions. The paper reports the main regulations and policies across Europe and the U.S., \minor{as well as} a few open challenges.} \\
      \hline
            2024 & \major{Satybaldy et al.} \cite{Satybaldy} & \major{A systematic literature review on SSI systems with particular emphasis on the main open challenges. It also reports a few available implementations.} \\
      \hline
      \textbf{\major{Now}} & {\major{\textbf{Our Survey}}}
      & A comprehensive survey on DIDs and VCs, \major{covering} all key aspects, \minor{including} \major{threats and mitigations,} primary implementations, applications across diverse domains, \major{and} emerging regulations/initiatives proposed by governments and organizations. \\
      \hline
    \end{tabular}
  \end{adjustbox}
\end{table*}

\begin{table*}[t!]
\centering
\caption{\major{Comparison of surveys in the field. \textbf{\major{Legend:}} 
\major{\CIRCLE \: Included, \LEFTcircle \: Partially Included, \Circle \: Not Included.}}}
\label{tab:comparison}
\begin{adjustbox}{width=\textwidth,center}
\begin{tabular}{c c c c c c c c}
\toprule
\textbf{\major{Year}} & \textbf{\major{Reference}} &\textbf{\makecell{\major{Literature}\\ \major{Review on SSI}}}&\textbf{\makecell{\major{Literature Review}\\ \major{on Other Applications}}} & \textbf{\makecell{\major{Threats \&}   \\ \major{Mitigation}}} & \textbf{\makecell{\major{Available}\\ \major{Implementations}}} & \textbf{\makecell{\major{Regulations, Projects}\\ \major{\& Organizations}}}& \textbf{\makecell{\major{Challenges \&} \\ \major{Future Directions}}}\\
\hline
\major{2021} & \major{Cucko et al. \cite{9558805}} &  \major{\Circle} & \major{\Circle} & \major{\Circle} & \major{\Circle}  & \major{\Circle} & \major{\LEFTcircle} \\
\major{2021} & \major{Soltani et al. \cite{Soltani2021}}   & \major{\CIRCLE} & \major{\LEFTcircle} & \Circle & \major{\LEFTcircle} & \major{\LEFTcircle} & \major{\LEFTcircle}\\
\major{2022} & \major{Bai et al. \cite{9881610}}  &\major{\Circle} & \major{\Circle} & \major{\Circle} & \major{\Circle} & \major{\Circle} & \major{\LEFTcircle}\\
\major{2022} & \major{Schardong et al. \cite{s22155641}}   & \major{\CIRCLE} & \major{\Circle} & \major{\Circle} & \major{\Circle} & \major{\Circle} & \major{\LEFTcircle}\\
\major{2023} & \major{Ernstberger et al. \cite{Ernstberger}}   & \major{\LEFTcircle} & \major{\Circle} & \major{\LEFTcircle} & \major{\CIRCLE} & \major{\Circle} & \major{\CIRCLE}\\
\major{2024} & \major{Krul et al. \cite{krul2024sok}}   & \major{\CIRCLE} & \major{\Circle} & \major{\CIRCLE} & \major{\CIRCLE} & \major{\Circle} & \major{\CIRCLE}\\
\major{2024} & \major{Tan et al. \cite{tan}}   & \major{\LEFTcircle} & \major{\LEFTcircle} & \major{\Circle} & \major{\Circle} & \major{\LEFTcircle} & \major{\LEFTcircle}\\
\major{2024} & \major{Satybaldy al. \cite{Satybaldy}}  & \major{\LEFTcircle} & \major{\LEFTcircle} & \major{\Circle} & \major{\LEFTcircle} & \major{\Circle} & \major{\CIRCLE}\\
\major{\textbf{Now}} & \major{\textbf{Our Survey}}  & \major{\CIRCLE} & \major{\CIRCLE} & \major{\CIRCLE} & \major{\CIRCLE} & \major{\CIRCLE} & \major{\CIRCLE} \\ 
\hline
\end{tabular}
\end{adjustbox}
\end{table*}

Consequently, such technologies can play a key role in establishing trust and secure communications among peers through digital identities \cite{9031548}, encompassing both human and non-human entities such as IoT devices. DIDs and VCs have been proposed as valuable solutions to enhance privacy and security in several application domains (e.g., smart transportation and smart healthcare). These standards can be extended to anyone and anything, including cloud, edge, and IoT resources. Notably, numerous implementations supporting these technologies have recently been developed and made available by different organizations, including major companies in the industry, such as Microsoft. Additionally, government institutions worldwide have been actively working to promote the widespread adoption of DIDs and VCs. \minor{For instance, in May 2024, the European Union introduced Regulation 2024/1183 \cite{eu2024regulation1183}, establishing the European Digital Identity Framework.} This initiative seeks to provide European citizens with secure access to online \minor{and offline public and private} services throughout Europe via an SSI system, reflecting a significant stride toward the advancement of digital identity solutions.

\subsection{Comparison with Related Surveys} 
Interest in DIDs and VCs has rapidly \major{grown} worldwide, \major{drawing significant} attention \major{from academia, industry, governments, and standardization bodies}. This trend is fueled by the \major{increasing} number of research papers and initiatives \major{emerging} in recent years. 
\major{However,} \major{most} existing surveys \major{primarily focus on the role of DIDs and VCs within SSI ecosystems, often overlooking their \minor{wider applications}. In contrast, this survey aims to provide a comprehensive overview of the use of these technologies. Table \ref{tab:comparison1} briefly summarizes key contributions of relevant surveys in the field, while Table \ref{tab:comparison} \minor{presents} a comparative analysis, highlighting their scope and focus. \texttt{INCLUDED} ($\CIRCLE$) means that the paper thoroughly covers the topic in question, while \texttt{PARTIALLY INCLUDED} ($\LEFTcircle$) indicates that the paper covers the topic to some extent but not in full depth.}

\smallskip
\noindent \textbf{\major{Literature Review on SSI and Other Applications.}} \major{This survey offers a comprehensive overview of how DIDs and VCs are employed across various areas, such as smart transportation and industry. Other surveys \minor{primarily focus } on SSI \cite{s22155641} or \minor{narrow their scope to} subtopics \cite{Soltani2021, krul2024sok, Satybaldy}. Beyond reviewing papers on SSI, Soltani et al. \cite{Soltani2021} also consider a few other application domains, such as healthcare and IoT. Tan et al. \cite{tan} cover a broader range of fields but their literature review does not emphasize how DIDs and VCs are used in these domains.}

\smallskip
\noindent \textbf{\major{Threats \& Mitigation.}}
\major{Threat and mitigation strategies are often overlooked in existing literature. Our survey is \minor{one of the few} to identify threats related to DIDs and VCs, \minor{while} also proposing potential mitigation strategies. Similar to our work, Krul et al. \cite{krul2024sok} discuss threats and mitigation in SSI systems, while Ernstberger et al. \cite{Ernstberger} present a concise threat model for SSI.}

\smallskip
\noindent \textbf{\major{Available Implementations.}} \major{This survey provides practical guidance for developers by identifying the most suitable solutions for their needs, unlike other works that primarily report frameworks and some of their features. Krul et al. \cite{krul2024sok} list the main industrial solutions and how they support identification, credential exchange, and some security and privacy properties.
Ernstberger et al. \cite{Ernstberger} offer tabular comparisons of supported functionalities, while other studies \minor{provide} a brief overview of a few frameworks \cite{Soltani2021, Satybaldy}.}

\smallskip
\noindent \textbf{\major{Regulations, Projects \& Organizations.}} \major{A key contribution of our work is an extensive review of regulations, projects, and organizations emerging worldwide. Only Tan et al. \cite{tan} and Soltani et al. \cite{Soltani2021} conduct similar research; however, their studies are limited to the main initiatives in Europe and the United States.}

\smallskip
\noindent \textbf{\major{Challenges \& Future Directions.}} \major{While almost all related works discuss challenges and future directions, most \minor{either address} them briefly or focus on sub-fields. For instance, Ernstberger et al. \cite{Ernstberger} and Krul et al. \cite{krul2024sok} focus on privacy and security challenges. Our survey identifies a wide range of challenges and future research directions specific to DIDs and VCs. Satybaldy et al. \cite{Satybaldy} provide the most comprehensive analysis of SSI challenges but they only present a few research directions.}

\begin{figure}[!t]
\centering
\resizebox{\columnwidth}{!}{\includegraphics{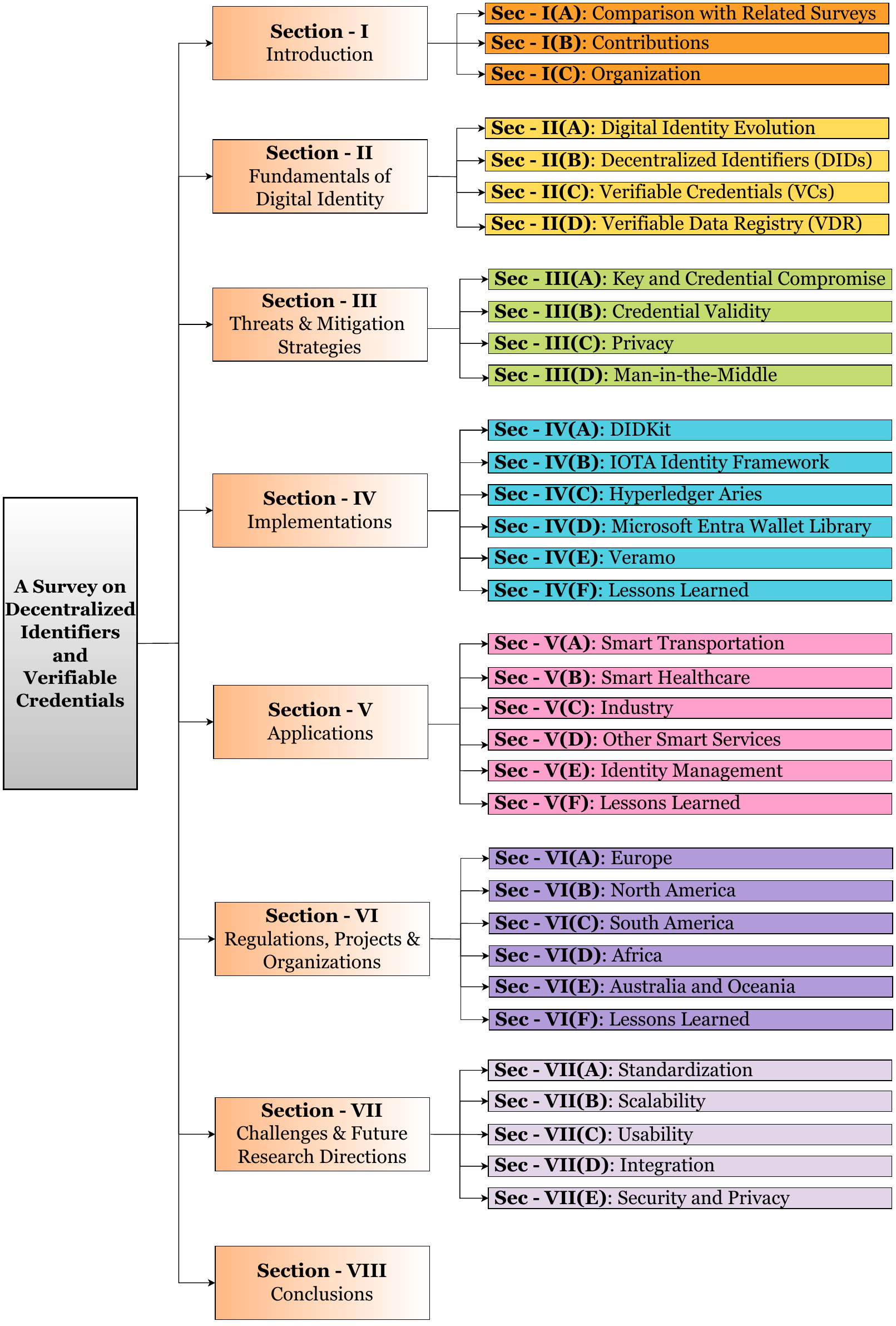}}
\caption{Illustrative organization of the survey.}
\label{Organization}
\end{figure}

\subsection{Contributions}

Related surveys \major{primarily focus on how} DIDs and VCs \major{have been employed} in SSI \major{systems}, often \major{overlooking critical areas such as security, practical} implementations, alternative applications, and regulations. 
\major{This} survey aims to bridge \major{these gaps} by \major{providing} an \major{in-depth examination of DIDs and VCs}. \major{It covers a wide range of topics, including foundational concepts, security threats, potential mitigation,} \minor{major} available implementations, application domains, \major{as well as challenges and emerging research directions}. We believe that this \major{paper} \major{offers a comprehensive resource for} readers \major{aiming to gain} thorough \major{understanding} of \major{these emerging technologies}. The key contributions of \major{our work} can be summarized as follows:

\begin{itemize}
    \item \major{We identify the key threats associated with DIDs and VCs and propose potential mitigation strategies to address them.}
    \item We \major{conduct a comparative analysis \minor{of} the main} commercial implementations available, aiding developers in \major{making informed decisions tailored} for their specific \major{needs}.
    \item We \major{review how} DIDs and VCs \major{have been employed} in research papers across diverse applications, \major{which goes beyond SSI systems}.
    \item We \major{report and discuss} a wide array of global initiatives, regulations, and projects that have emerged, providing valuable insights into \major{the potential of these technologies and how they have been adopted worldwide.}
    \item We identify and highlight critical challenges that \major{are specific} to DIDs and VCs. \minor{Moreover}, we \major{point out under-explored areas that warrant further investigation}, outlining promising research directions. 
\end{itemize}

\begin{figure*}[!t]
\centering
\includegraphics[width=0.85\textwidth]{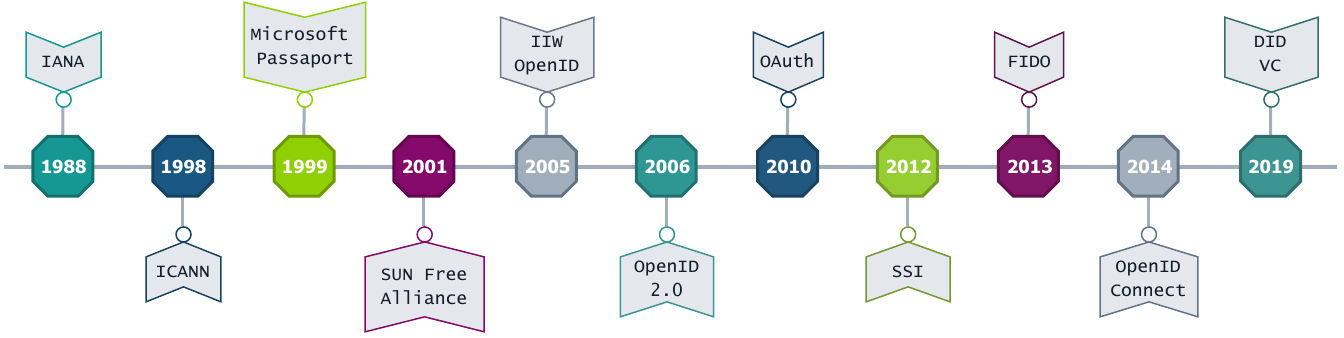}
\caption{Timeline of digital identity evolution.}
\label{digitalevolution}
\end{figure*}

\subsection{Organization}
Figure \ref{Organization} \major{illustrates the structure of this survey}, the remainder of this paper is organized as follows. Section \ref{background} \minor{provides} the \minor{foundation of digital identities by} outlining their evolution over \minor{time}, with particular \minor{focus} on DIDs and VCs.  \major{Section \ref{sec:threats} identifies key security threats \minor{associated with DIDs and VCs. It also} discusses potential mitigation strategies,} \minor{which are crucial for ensuring the secure adoption of these technologies.}
Section \ref{implementation} \minor{shifts from theoretical aspects to practical considerations, exploring the spectrum of} main available implementations. \minor{This section helps} developers in selecting the most suitable solution based on their specific \minor{needs}. Section \ref{application} \minor{links the technical and practical aspects by reviewing} the various applications of DIDs and VCs across different research \minor{domains, demonstrating their real-world impact and versatility.}
Section \ref{regulation} \minor{examines the broader ecosystem by offering} insights into global regulations, projects, and organizations.
Section \ref{future} \minor{identifies} existing challenges that \minor{hinder the widespread} adoption of these technologies \minor{and highlights opportunities for} future research, \minor{bridging current limitations with potential advancements.} 
Finally, Section \ref{conclusion} \minor{summarizes the key insights from the survey and provides concluding remarks.} Table \ref{tab:acronyms} \minor{reports} a list of abbreviations frequently used \minor{throughout the paper}.

\begin{table}
    \centering
    \caption{List of the acronyms frequently used in this article.}
\label{tab:acronyms}
\begin{tabular}{ c c}
\hline
\textbf{Acronym} & \textbf{Description} \\
\hline
ABAC & Attribute-based Access Control\\
CID & Content Identifier \\
DAG & Direct Acyclic Graph \\
DID & Decentralized Identifier \\ 
DHS & Department of Homeland Security\\
DIACC &Digital Identity and Authentication Council of Canada\\
DLT & Distributed Ledger Technology \\ 
DT & Digital Twin\\
eIDAS &  Electronic Identification and Trust Services \\
EBSI & European Blockchain Services Infrastructure\\
EU & European Union\\
EUDIW & European Digital Wallet\\
EV & Electric Vehicle \\
FL & Federated Learning \\
FIDO & Fast IDentity Online\\
5G & 5th Generation Mobile Networks\\
GDPR & General Data Protection Regulation\\
IIoT & Industrial Internet of Things\\
IoMT & Internet of Medical Things\\
IoT & Internet of Things\\
IoV & Internet of Vehicles\\
IPFS & Internet Planetary File System\\
ML & Machine Learning\\
OAuth & Open Authorization\\
OEM & Original Equipment Manufacturer\\
SSI & Self-Sovereign Identity \\
SSO & Single-Sign On\\
U.S. & United States\\
VC & Verifiable Credential \\
VDR & Verifiable Data Registry \\
VIN & Vehicle Identification Number\\
V2X & Vehicle-to-anything \\
V2V & Vehicle-to-vehicle \\
VP & Verifiable Presentation \\
W3C & World Wide Web \minor{Consortium} \\
ZKP & Zero-Knowledge Proof\\
\hline
\end{tabular}

\end{table}

\section{Fundamentals of Digital Identity}\label{background}
This section \minor{provides} an overview of the development of digital identities, \minor{with a focus} on the standards and technologies that enable SSI systems. 

\subsection{Digital Identity Evolution}
\major{The evolution of} digital identity has \major{undergone} many eras, which have gradually shifted digital identification from centralized to decentralized \minor{identity} models \cite{ANTE2022523}. Figure \ref{digitalevolution} \minor{depicts} \minor{this} timeline evolution.

\smallskip
\noindent \textbf{Centralized Identity.}
In centralized identity systems, \minor{identities are managed by a central authority} \cite{10.1145/3407023.3407026}. 
One of the \major{earliest} forms of digital identity dates back to 1988 when the Internet Assigned Numbers Authority (IANA) was responsible for determining the validity of IP \minor{addresses}. Later, the scientific community witnessed the \minor{emergence} of organizations such as the Internet Corporation for Assigned Names and Numbers (ICANN), \minor{which arbitrates} the validity of domain names. 

\major{Centralized identity management systems rely on usernames and passwords for authentication, which raises several concerns \cite{Yild2019}, as} users \minor{often prioritize} simplicity over security when choosing their \minor{credentials. This makes} them vulnerable to attacks \minor{such as} dictionary or phishing attacks. \major{Moreover,} individuals \major{usually} have as many identities as the number of services, \minor{resulting in a} fragmented \minor{identity} landscape.

\major{Centralized models} suffer from a single point of failure and \major{expose user-related} information to potential \minor{risks, whether from} centralized authorities or through data breaches \cite{saleem2020sok, peteretal}. Consequently, this identity model \major{also} \minor{poses challenges} for service providers \cite{7980332}, \minor{requiring} substantial investments \minor{to} securely \minor{store, preserve, and safeguard} user data in compliance with existing regulations.

\smallskip
\noindent \textbf{Federated Identity.}
Federated identity represents \minor{the} second era of digital identification, \major{addressing} some \minor{of the challenges associated with} centralized identification. \minor{It} allows users to \minor{utilize} the same identity across multiple sites and applications \cite{Chadwick2009}. By logging in once, users can access various resources without creating separate accounts for each system. This approach relies on trusted relationships between different parties \minor{to ensure} secure authentication and authorization. Indeed, \minor{federations are} achieved by distributing identification and verification components across systems or \minor{by} mutually accepting \minor{shared} standards \minor{between them}.

One of the pioneers in proposing federated identity was Microsoft with its Passport program, \minor{which allowed} users to access different websites \major{with a single login}. In 2001, Sun Microsystems formed the Liberty Alliance to develop open standards for federated identity and identity-based web services. \major{Today}, major companies like Google and Meta \minor{offer} Single Sign-On (SSO) support \cite{10.1145/1456396.1456397}. A typical federated scenario involves a user authenticating \minor{through} an identity provider (e.g., Google or Meta identity providers) \minor{and then gaining access to} other applications on its behalf. 

While federated models \major{reduce} identity fragmentation, they still pose privacy \minor{concerns.} Despite a decrease in the number of centralized entities involved, users must trust identity providers, which remain centralized, \minor{offering limited control over user} data. Furthermore, federated identity relies on mutual recognition between two or more parties, which may be \minor{increasingly} complex at scale \cite{7395570}.

\begin{figure}[!t]
\centering
\includegraphics[width=0.49\textwidth]{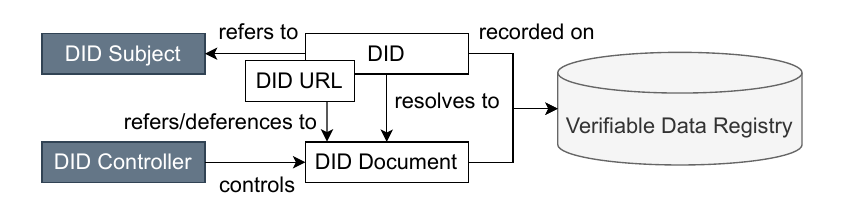}
\caption{Overview of a DID-based architecture and the relationship of the main components.}
\label{didarch}
\end{figure}

\smallskip
\noindent \textbf{User-centric Identity.}
The demand for \major{greater} control over personal data has given rise to user-centric identity. This model \minor{empowers} individuals to manage their identities \minor{independently} or across multiple authorities without relying on centralized federations \cite{9045440, 10.1145/1179529.1179532}. Users retain autonomy and must explicitly grant consent before sharing or modifying their data, enhancing privacy and security. 

In 2005, the Identity Commons, an influential American organization with a mission to support, facilitate, and promote the creation of an open identity layer for the Internet, played a pivotal role in advocating for the Internet Identity Workshop (IIW). The IIW places users at the core of identity management, aiming to empower individuals in shaping their online identities. \major{This lead to} numerous projects \minor{embodying} user-centric principles, including OpenID, OpenID 2.0, OpenID Connect (OIDC), Open Authorization (OAuth), and Fast IDentity Online (FIDO) \cite{srinivas2015universal, 10.1145/1179529.1179532, hardt2012oauth}. User-centric approaches \major{enable} individuals to control their data through various authenticators (e.g., JSON Web Tokens) and certificates issued by different service providers, all securely stored \minor{on} their devices. 

Despite the notable progress, user-centric identity systems still depend on trust relationships with service providers and authorities, \major{requiring} users \major{to} trust these entities \major{to} handle their information responsibly and not misuse it \cite{Jøsang2015}.

\smallskip
\noindent \textbf{Self-Sovereign Identity.}
SSI represents the last era \major{in the} long and evolving journey of digital identity \major{evolution}. \major{Introduced in 2012, this} concept \minor{marks} a significant advancement over user-centric identity systems \cite{tobin2016inevitable}. 

In contrast to conventional identity management systems, where the service provider \minor{controls} the identity model, SSI places the individual at the core. Users are granted full control over their identities, enabling them to decide when, if, and how they wish to disclose or modify their data. SSI \minor{relies on} decentralized infrastructure and cutting-edge technologies such as DLTs, DIDs, and VCs. By leveraging DLTs like \textit{blockchain}, SSI eliminates the need for central authorities \cite{cao_survey}, \major{fostering} a trustless environment where users can interact with services and applications without \minor{exposing} their sensitive information to a single entity \cite{8776589}. This decentralized architecture ensures that data is cryptographically secured and verifiable, \minor{delivering} transparency and immutability for identity management. 

In an SSI ecosystem, individuals are uniquely identified by DID, linked to a key pair \minor{controlled by} the user. The public key bound to the identifier is \major{usually} shared on a DLT, \major{allowing users to verify identities independently} \cite{MUHLE201880}. Users prove attributes or claims through VCs, \major{which} can \minor{be} cryptographically verified by any other parties without \minor{interacting with the issuing} centralized authority. \minor{These credentials} are typically transmitted off-chain due to privacy considerations, with verification \minor{depending} on publicly available information \minor{associated with} the identifier and credential issuer.

\begin{figure}[!t]
\centering
\includegraphics[width=0.45\textwidth]{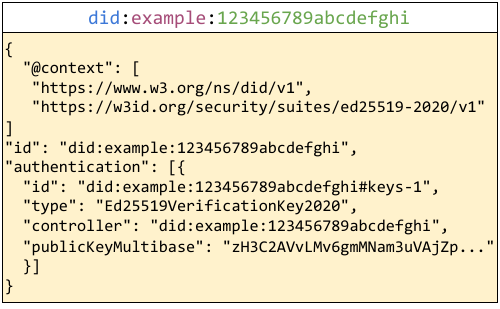}
\caption{An example of a DID and its DID Document.}
\label{did}
\end{figure}

\subsection{Decentralized Identifiers (DIDs)}
DIDs have emerged as a groundbreaking \major{digital} identifier widely embraced within decentralized identity systems. They were formalized and standardized by the World Wide Web Consortium (W3C) \cite{w3c} \major{after} substantial collaborative efforts from 2017 to 2019, culminating in the publication of the DID specification as an official W3C Recommendation. Figure \ref{didarch} \major{provides an} overview of the major components of a DID-based architecture.

\smallskip
\noindent \textbf{\major{Architecture.}} A DID uniquely identifies a DID Subject, \major{which can be either} a human or non-human entity. It \major{consists of} three essential components: the Uniform Resource Identifier, the identifier for the specific DID method, and the method-specific identifier for the DID. \major{The} DID method specifies the \major{processes for creating, resolving, updating, and deactivating DIDs} and DID Documents. \major{The} DID URL \major{extends} a basic DID by including additional URI components like path, query, and fragment, \major{enabling} precise resource location within a DID Document or an external resource. 

Each DID resolves to a DID Document, a machine-readable JSON-LD document \major{containing} information about the DID Subject, \major{such as} cryptographic public keys, service endpoints, authentication parameters, timestamps, and additional metadata. \major{DIDs are consistent and permanent, offering reliable identification even as individuals switch service providers or platforms.} Figure \ref{did} shows an example of DID and its corresponding DID Document.

DIDs \major{are} designed to \major{eliminate reliance on centralized identity providers, fostering the adoption of SSI systems. Users have control and ownership over their DIDs, which can be created and managed by themselves.} An entity can prove the ownership of a DID by leveraging the private key corresponding to the public \major{key} in the DID document. \major{The verifier can access DID Document}, which is publicly available through a Verifiable Data Registry (VDR). 

The DID Controller \major{is} the entity \major{authorized} to modify the DID Document. A DID might have \major{multiple controllers} or \major{may} coincide with the DID Subject, \minor{aligning with} the principles of the SSI paradigm. DIDs are resolved through a universal resolver \cite{sabadello2017universal} that supports multiple DID systems. A DID system \major{needs to implement} a DID adapter \major{to be compatible with the universal resolver,} \minor{functioning}  as an interface between system-specific DID methods and the universal resolver.

\smallskip
\noindent \textbf{\major{Types of DIDs.}} The DID specification \cite{didspec} introduces three types of DIDs:

\begin{itemize}
    \item Anywise DIDs can be utilized with an unspecified number of parties, typically strangers. \minor{They allow} broad usage without limiting the number of relationships that can be established.
    \item Pairwise DIDs \minor{are} only known by \minor{their} subject and one other party, such as a service provider. Pairwise DIDs address privacy concerns by ensuring that each relationship has a unique DID, minimizing the risk of correlation between different interactions.
    \item N-wise DIDs \minor{are} designed to be known by strictly N parties, including the subject. \minor{They} encompass pairwise DIDs as a special case when N equals 2.
\end{itemize}

\begin{figure}[!t]
\centering
\includegraphics[width=0.49\textwidth]{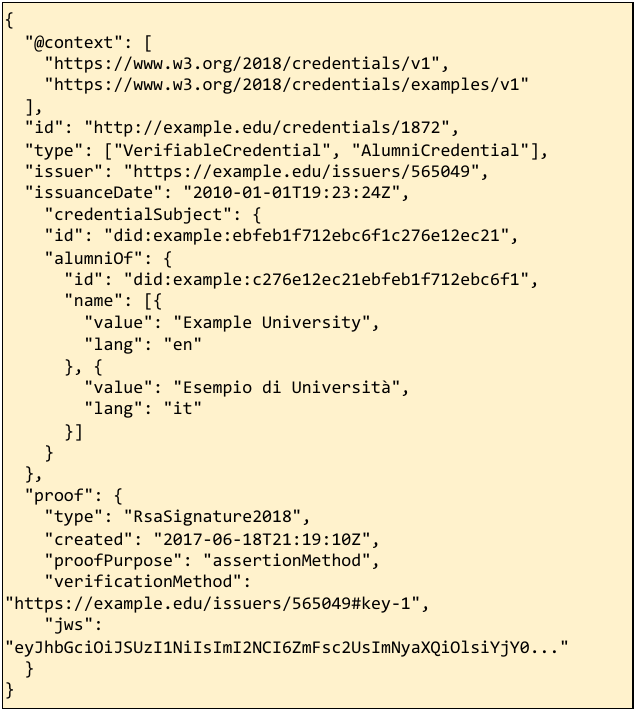}
\caption{\minor{Example of a VC allowing alumni of Example University to receive discounts on season tickets for sporting events.}}
\label{vcex}
\end{figure}

\smallskip
\noindent \textbf{\major{Communication Protocols.}} The growing interest in DID has also \major{led to the development} of DID-based communication protocols \cite{a16010004, sabadello2018introduction}, \major{enabling} private and secure communication between two or more SSI entities. \minor{These protocols} rely on DIDs and facilitate mutual authentication between the participating parties. 

The DID Auth protocol allows an identity owner to use their client application, such as a mobile device or browser, to prove their control over a DID to a service provider. This protocol utilizes a challenge-response cycle that can be customized based on the specific \major{circumstance, potentially replacing} traditional forms authentication \major{methods} like usernames and passwords, \major{establishing} an authenticated communication channel between the identity owner and the service provider. 

\subsection{Verifiable Credentials (VCs)}
VC is \major{a} specification developed by the W3C \major{to} create an interoperable data structure capable of representing claims \major{(e.g., properties or attributes)} that are cryptographically verifiable and tamper-proof. VCs are designed to seamlessly operate across different platforms and applications. Individuals can store VCs in their digital wallet and \minor{carry} them anywhere while \minor{ensuring they remain} verifiable. \minor{This} flexibility and convenience \minor{allow users to present} their credentials \minor{efficiently}. Figure \ref{vcex} shows an example of VC that enables all alumni of "Example University" to \major{receive} a discount on season tickets to sporting events.  

\begin{figure}[!t]
\centering
\includegraphics[width=0.49\textwidth]{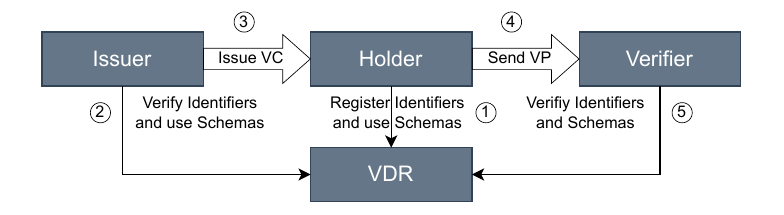}
\caption{Overview of VC main actors.}
\label{vcarch}
\end{figure}

\smallskip
\noindent \textbf{\major{Main Actors.}} In the VC ecosystem, a holder refers to an entity exercising control over one or more VCs. \major{These credentials are issued by trusted entities such as government agencies or banks.} A verifier is an entity, \major{such as} an e-commerce website, that requires valid credentials to offer a service. Figure \ref{vcarch} illustrates the key roles within the ecosystem of VCs.

To facilitate the creation and verification of identifiers, keys, verifiable credentials schema, and other relevant data essential for using VCs, \major{are shared through a} VDR, \major{which} acts as a mediator within the ecosystem. \major{Since credentials often contain sensitive information, they are typically shared off-chain instead of being stored on a VDR or a centralized system. This \minor{approach} minimizes the risk of data exposure by keeping sensitive information out of public or easily accessible systems, ensuring greater privacy and security for the user.}

\smallskip
\noindent \textbf{\major{Structure of VCs.}}
\major{Similar to} DIDs, VCs also consist of several elements, \minor{including} the Subject URI, the URI of the issuer responsible for the claims, and URIs that uniquely identify the credential. The \major{Subject} URI \major{is used to retrieve the} subject's public key, ensuring verification of the credential's ownership. \minor{Meanwhile,} the issuer URI is essential for obtaining their public key and verifying that the credential originates from a trusted entity. Notably, the URI can also take the form of a DID.  

Furthermore, VCs include claim expiration conditions and cryptographic signatures, \minor{ensuring} that the credential validity is maintained over time and that the integrity of the information is secured through cryptographic verification.

\begin{figure}[!t]
\centering
\includegraphics[width=0.49\textwidth]{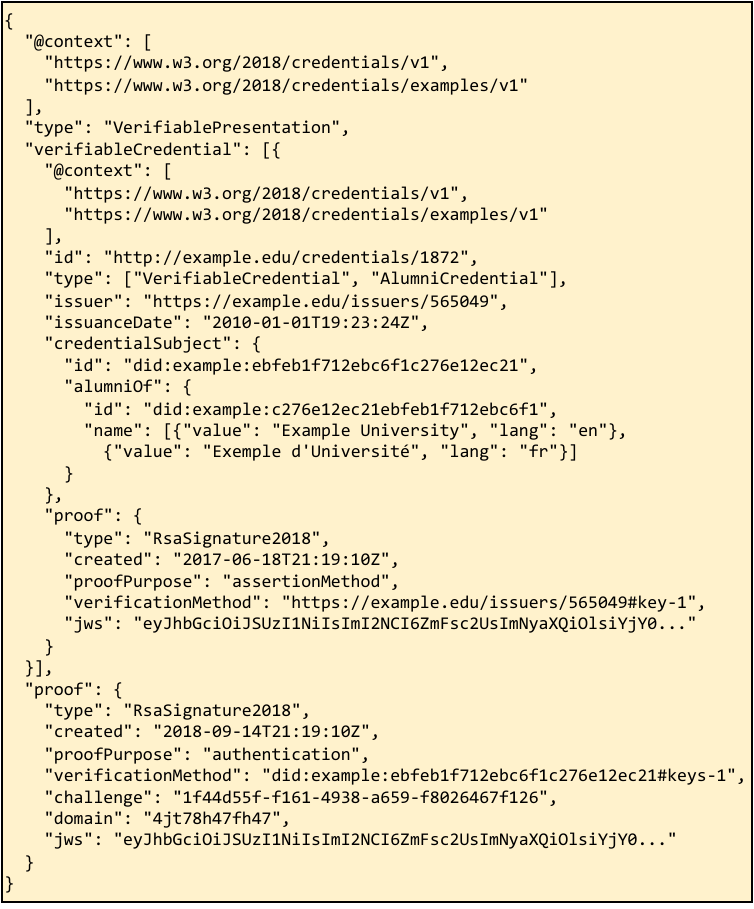}
\caption{\minor{Example of a VP derived from the VC for alumni of Example University.}}
\label{vpex}
\end{figure}

\smallskip
\noindent \textbf{\major{Verifiable Presentation.}}
The W3C Verifiable Credentials Working Group has also defined the concept of Verifiable Presentations (VPs), which specify the methods for signing and presenting VCs by the holder. VCs or VPs can be described using JSON-LD, JSON, or JSON Web Token formats. Figure \ref{vpex} shows \minor{a} VP obtained from the VC of the previous example. The standard related to expressing verifiable information on the Web has been published to provide a regulation specification for expressing credentials on the Internet \minor{ensuring that} they are machine-verifiable, respect privacy, and, most importantly, are cryptographically secure.

\smallskip
\noindent \textbf{\major{Selective Disclosure.}}
\major{To grant individuals full control over their data, VCs allow selectively disclosing a subset of the information contained \minor{with them}. Over the years, various methods for selective disclosure have been developed \cite{surveysd, FLAMINI2024103789}. \minor{They are usually} categorized into mono claims, hashed values, Zero-Knowledge Proofs (ZKP), and selective disclosure signatures. The current state-of-the-art solution is SD-JWT \cite{SD-JWT}, which enhances privacy by replacing plaintext claims with digests of their salted values. When the holder discloses specific information, they share the original claim and the corresponding salt. 
}

\subsection{Verifiable Data Registry (VDR)}
In the context of DIDs and VCs, a VDR plays a crucial role in \major{creating, managing, and verifying} identifiers, keys, credential schema, and other relevant data required to utilize these decentralized identity technologies \cite{10089278}. 

\smallskip
\noindent \textbf{\major{Role of VDR in SSI.}}
A VDR acts as a trusted intermediary within the ecosystem of DIDs and VCs, \major{serving} as a repository or database that stores and provides access to essential information. This includes DID Documents, which contain \major{details} such as cryptographic public keys, service endpoints, authentication parameters, timestamps, and metadata. The VDR ensures the availability and accessibility of these \major{documents} to support the resolution and verification processes. When verifiers need to verify a DID or a VC, they retrieve the relevant information from the VDR. By accessing the public information stored in the registry, \minor{they} can validate the authenticity, integrity, and validity of the DIDs and VCs involved in the verification process, \major{thereby establishing} trust and confidence in the decentralized identity ecosystem. 

The VDR also \major{manages and maintains} the lifecycle of DIDs and VCs. It facilitates the creation, registration, and revocation of DIDs and VCs, acting as a centralized point of control or coordination. For example, when an entity wants to create a new DID or issue a new VC, it interacts with the VDR to ensure proper registration and management of the associated information. Similarly, when a DID or VC needs to be revoked or updated, the VDR can handle the necessary operations to reflect the changes in the system. Additionally, the VDR supports interoperability and standardization within the decentralized identity ecosystem \major{by enforcing} consistent data formats, validation rules, and data-sharing protocols. This promotes compatibility and seamless integration across different DID methods, VC issuers, and verifiers.

\smallskip
\noindent \textbf{\major{Implementations.}}
\major{Currently,} the W3C's standards do \minor{not} specify how the VDR \major{should be implemented}. Most of the proposed approaches are based on DLTs, with blockchain standing out as the most popular type of DLT \cite{10.1145/3465481.3469204}. \major{Blockchain} is characterized by interconnected blocks, each \major{referencing} to the previous block hash, \major{forming} a secure chain \cite{8425607}. Any tampering would lead to a different hash that can be easily detected. This feature is fundamental \major{for} the secure sharing of information such as DID Documents. Additionally, many blockchains also support \textit{smart contracts} \cite{8847638}, which are self-executing programs stored directly on the blockchain and triggered when specific conditions are met. This inherent design ensures fault tolerance, tamper-proofing, and traceability. In the subsequent sections, we will show how these features make smart contracts a valuable technology in various application domains utilizing DIDs and VCs.

\major{Alternative solutions have also emerged}. For example, Information-Centric Networking (ICN) is a novel networking paradigm that can be adopted to implement the registry \cite{9146840}. Users interact with edge nodes to manage DIDs through HTTP APIs, which are then translated into appropriate ICN flows. 

\section{\major{Threats \& Mitigation Strategies}}\label{sec:threats}

\major{
Although DIDs and VCs offer significant security and privacy benefits, they are not without vulnerabilities. This section outlines \minor{the} major threats associated with these technologies and presents potential strategies \minor{for mitigation}.}

\subsection{\major{Key and Credential Compromise}}

\smallskip
\noindent \textbf{\major{Threat 1}}: \major{An adversary may perform various attacks, such as phishing, malware, or direct key theft, to compromise the private key associated with a DID or gain unauthorized access to credentials, impersonating the legitimate identity owner.}

\smallskip 
\noindent \textbf{\major{Mitigation:}}
\major{Mitigating the risks of key and credential compromise requires a multi-faceted approach. Below are some effective strategies:}

\begin{itemize}
    \item \textit{\major{Key Rotation}}: \major{Regularly rotating keys and revoking old ones minimizes the damage caused by a potential compromise.}
    \item \textit{\major{Multi-Factor Authentication (MFA)}}: \major{Implementing MFA enhances the security of wallet access by protecting keys and credentials. MFA can also require the presentation of the same credential from multiple devices, ensuring that compromising a single wallet does not allow the adversary to use the credential as their own.} 
    \item \textit{\major{Hardware Security Modules (HSMs)}}: \minor{Storing} private keys in secure hardware modules \minor{protects} them from unauthorized access.
    \item \textit{\major{Trusted Third Party Key Recovery}}: \major{Keys can be securely backed up and recovered through trusted third parties \cite{8890476}, using escrow services to safeguard against key loss.}
\end{itemize}
\smallskip

\noindent \textbf{\major{Threat 2}}: A malicious user may collude \minor{with others} or steal a valid credential from a legitimate user, presenting it to a verifier and acting on behalf of the holder.

\smallskip 
\noindent \textbf{\major{Mitigation:}} This \minor{threat} can be mitigated by associating the VC and their presentation \minor{with} the identity owner, preventing illegitimate transfers. The holder identifier, such as a DID, is included in the VC, whose immutability prevents any \minor{alteration}. To prove ownership over the credential, the presenter must demonstrate knowledge of the identity \minor{owner's} private key, which is only known by the holder.

\smallskip
\noindent \textbf{\major{Threat 3}}: An attacker may attempt to gain access to a service by forging a legitimate VC or presenting a credential issued by an entity that \minor{lacks} the authority to issue VCs.

\smallskip 
\noindent \textbf{\major{Mitigation:}} Credentials must be digitally signed by the issuer, \minor{preventing} adversaries from forging VCs as they do not know the \minor{issuer's} private key. Each credential must include an issuer identifier \minor{to} ensure that a trusted and recognized authority has certified the presented claims. \minor{Additionally,} a secure register \minor{should be maintained to identify} all trustworthy issuers and \minor{the} claims they \minor{are authorized to} certify.

\begin{figure}[!t]
\centering
\includegraphics[width=0.49\textwidth]{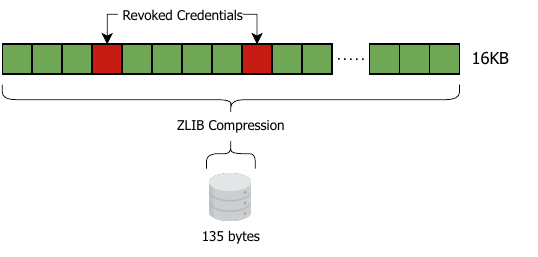}
\caption{Revocation List 2020.}
\label{revocation}
\end{figure}

\subsection{\major{Credential Validity}}

\noindent \textbf{\major{Threat 4}}:
Although always verifiable, the validity of VCs can change over time due to loss of privileges or expiration. Consequently, a verifier must verify the validity of a credential before \minor{accepting it.}

\smallskip 
\noindent \textbf{\major{Mitigation:}}
The issuer can specify the validity period within the credential itself, \minor{enabling} the verifier to check whether the credential is still valid.

However, this approach does not \minor{account for scenarios where} issuers revoke VCs due to \minor{a holder's} loss of privileges. This concern can be \minor{addressed using} revocation mechanisms \minor{commonly employed} for PKI certificates \cite{10285344}, such as Online Certificate Status Protocol (OCSP) \cite{santesson2013x} and Certificate Revocation Lists (CRLs) \cite{cooper2008internet}. The W3C also proposed the revocation bitstring "Revocation List 2020" \cite{rl}, where each bit corresponds to an index contained in the VCs. Figure \ref{revocation} \minor{provides} an illustrative overview of Revocation List 2020. When \minor{a} bit is set to 1, the \minor{corresponding} VC is revoked; otherwise, it \minor{remains} valid. \minor{Since most} credentials \minor{are expected to remain} valid, \minor{long} stretches of bits often retain identical values. This \minor{property makes the bitstring amenable} to compression techniques like ZLIB\cite{gailly2004zlib}, \minor{which can reduce its size. Furthermore,} concerns \minor{about the size of} data structures for managing revocation have \minor{inspired} novel approaches that minimize storage and network requirements \cite{mazzoccausenix}.  

\begin{table*}[t!]
    \caption{Comparative analysis of main implementations of DIDs and VCs. \major{\textbf{Legend:} \textbf{$\uparrow$} High, \textbf{$\approx$} Medium, \textbf{$\downarrow$} Low, \textbf{-} Not Specified.}}
\label{tab:implementations}
\begin{adjustbox}{width=\textwidth,center}
\begin{tabular}{l c c c c c c c}

\hline
\textbf{Library} & \textbf{DIDKit} & \textbf{\makecell{IOTA Identity\\ Framework}} & \textbf{\makecell{Hyperledger\\ Aries}} & \textbf{\makecell{Microsoft Entra\\Wallet}} & \textbf{Veramo} \\
\hline
\textbf{Main Target Platform} & Multi-platforms & Multi-platforms/IoT Devices & Multi-platforms & Mobile Applications & Multi-platforms \\
\hline
\textbf{Programming Language} & \makecell{Rust, C, Java, Android,\\ Python, JavaScript} & Rust, Node.js & Python, JavaScript, Go, .NET & Android, iOS & JavaScript \\
\hline
\textbf{\major{W3C Compliance}} & \checkmark & \checkmark & \checkmark & \checkmark & \checkmark \\
\hline
\textbf{\major{Credential Format}} & \major{JSON-LD/JWT} & \major{JWT} & \major{JSON-LD} & \major{JWT} & \major{JSON-LD/JWT} \\
\hline
\textbf{\major{Key \& Wallet Management}} & \major{\textbf{-}} & \major{Stronghold} & \major{\textbf{-}} & \major{Azure Key Vault} & \major{\textbf{-}} \\
\hline
\textbf{\major{Verifiable Proof Types}} & \makecell{\major{RSA/EdDSA/ECDSA/}\\ \major{EIP712/JWS2020}} & \major{EdDSA/ECDSA} & \major{BBS+/EdDSA} & \major{EdDSA/ES256K/ECDSA P-256} & \major{EdDSA, ECDH, ECDSA} \\
\hline
\textbf{\major{Selective Disclosure}} & \major{SD-JWT} & \major{SD-JWT/ZKSD} & \major{SD-JWT} & \major{SD-JWT} & \makecell{\major{SD-JWT Through}\\ \major{Plugin Interface}} \\
\hline
\textbf{\major{Verifiable Data Registry}} & \major{\textbf{-}} & \major{Tangle} & \major{Indy Ledger} & \major{\textbf{-}} & \major{\textbf{-}} \\
\hline
\major{\textbf{Learning Curve}} & \major{$\downarrow$} & \major{$\approx$} & \major{$\uparrow$} & \major{$\approx$}  & \major{$\downarrow$} \\
\hline
\textbf{Open Source} & \checkmark & \checkmark & \checkmark & \checkmark & \checkmark \\
\hline
\end{tabular}
\end{adjustbox}
\end{table*}

\subsection{\major{Privacy}}

\smallskip
\noindent \textbf{\major{Threat 5}}:
A VC may contain more claims than necessary for accessing a specific service. Consequently, a service provider \minor{might} acquire more information than required. Malicious service providers \minor{could} exploit personal data for financial gain \minor{through} individual profiling, \minor{while} honest service providers \minor{might inadvertently} elevate privacy loss risks.

\smallskip 
\noindent \textbf{\major{Mitigation:}}
\minor{Individuals} should provide only the information needed to access the intended service \minor{by} removing unnecessary claims from a VC when generating the presentation. This can be achieved \minor{through} employing selective disclosure \cite{surveysd, FLAMINI2024103789}, which allows presenting only a subset of claims contained in a VC. Selective disclosure techniques enable verifying the authenticity of the revealed claims without accessing the entire credential.

\smallskip
\noindent \textbf{\major{Threat 6}}:
Selective disclosure techniques could still allow service providers to link claims to the same individual across subsequent presentations or collude with other providers \cite{krul2024sok}.

\smallskip 
\noindent \textbf{\major{Mitigation:}}
\major{Using Pairwise DIDs can help mitigate this risk by reducing the likelihood of correlation, \minor{though} it does not entirely prevent linkage across multiple presentations. \minor{To further reduce the risk}, VCs should avoid including persistent user identifiers, such as DIDs, \minor{unless} necessary for issuer identification. While \textit{complete unlinkability} \minor{might} be ideal in some scenarios, it is not always appropriate. For instance, when a credential is used for authentication, the service provider must recognize repeat presentations to prevent Sybil attacks. \minor{Linked unlinkability} approaches \cite{zhang2021passo} \minor{can address this}, \minor{allowing} the identification of repeat presentations while minimizing the risk of correlation \minor{across different} providers.}

\noindent \textbf{\major{Threat 7}}:
\major{The holder of a VC may not have permission to access certain confidential information within the credential, which is intended \minor{solely} for a verifier.}

\smallskip 
\noindent \textbf{\major{Mitigation:}}
\major{To protect sensitive information, claims within the VC can be encrypted using a secret key known only to authorized service providers. Additionally, selective disclosure mechanisms can be employed, \minor{ensuring} only the necessary claims to be revealed. In such cases, the information required to reveal claims is shared \minor{exclusively} with the service providers, not the holder.}

\subsection{\major{Man-in-the-Middle}}

\noindent \textbf{\major{Threat 8}}:
\major{An attacker could intercept and tamper with the communication between a legitimate user and a service provider, potentially gaining access to the contents of the presentation or altering the transmitted information.}

\smallskip 
\noindent \textbf{\major{Mitigation:}}
Communications between two parties can be secured by using end-to-end encryption protocols \cite{10.1145/3580522}, \minor{which protect} the content from unauthorized access. \minor{Additionally,} as VCs are signed with the issuer's private key, \minor{any modification of} their content \minor{would be} detected.

\noindent \textbf{\major{Threat 9}}:
\major{An adversary may intercept a VP and reuse it to a different verifier, impersonating the legitimate holder and gaining unauthorized access.}

\smallskip 
\noindent \textbf{\major{Mitigation:}}
\major{Replay attacks can be mitigated by including a nonce or a timestamp in the VP. Verifiers should send a challenge set as the nonce within the credential. To be valid, the VP must include the challenge, \minor{ensuring it} matches \minor{the one} provided by the verifier.}

\section{Implementations}\label{implementation}
The successful implementation of DIDs and VCs plays a crucial role \minor{in} the \major{widespread} adoption of SSI systems and the development of next-generation services. Developers \major{need} efficient \major{frameworks} that \major{simplify the use of} these technologies while adhering to \major{established} \minor{standards}. 

\major{Although initially designed to give individuals greater} control over personal data, \major{DIDs and VCs have broader applicability, extending to entities} \minor{such as} cloud, edge, and IoT devices, which \major{often} feature different storage and computational capabilities. \major{This heterogeneity} demands implementations that \major{are optimized to minimize storage and computational overhead.} 

The development of DIDs and VCs \major{remains} in its early stages, \major{with a relatively small number of} available implementations. \major{Nonetheless}, our \major{analysis provides} developers and stakeholders \major{with} valuable insights into the key differences among the main solutions, guiding them \minor{in choosing} the most suitable \minor{option} for their \major{specific} use cases. Table \ref{tab:implementations} \major{reports} the key features of main implementations. 

\subsection{DIDKit}
DIDKit \cite{didkit} is a toolkit from SpruceID that \major{provides} DID and VC functionalities across \minor{various} platforms. \major{Its} core libraries are \major{implemented} in Rust, \major{offering advantages} \minor{such as} memory safety, \major{simpler dependency tree,} and \major{compatibility with} various platforms, including embedded systems. 

\major{While } DIDKit SDK \major{supports a wide range of use cases, its bindings for other languages—such as C, Java, Android, Python, and JavaScript—are implemented as wrappers around the Rust core. As a result, using these bindings may \minor{introduce} additional challenges and complexity compared to direct Rust usage. \minor{DIDKit} supports selective disclosure, implemented through SD-JWT.} \minor{This library} offers the following features:

\begin{enumerate}
    \item It can sign and verify any VC compliant to the \major{W3C} standard. DIDKit also includes a \minor{ready-to-use} HTTP/HTTPS server that \minor{accessed via} any API interface, including those specified by the W3C standard.  
    \item It \minor{supports} and translates both \minor{of} the two major signing systems and proof formats used in VC.
    \item It can handle, authenticate, validate, register, and deterministically generate many kinds of DIDs compliant with the W3C standard.
    \item It can also issue and consume authorization tokens based on the Object Capabilities "ZCaps".
    
\end{enumerate}

\subsection{IOTA Identity Framework}
The IOTA Identity framework \cite{iotaidentity} implements \major{decentralized identity solutions using both} a DLT-agnostic \minor{approach} and \major{dedicated IOTA method specification}. It is \major{built on} the Tangle, a next-generation DLT \major{tailored} for the IoT \minor{ecosystem}. The IOTA Identity Framework enables the creation of new digital identities for \major{any entity}, including IoT devices, at any time. \major{These} functionalities are offered in Rust and \major{Node.js via} Web Assembly (WASM), making it versatile across various environments. 

When using IOTA Identity, the verifier can confirm the issuer's identity through their public key on the Tangle, while the holder provides proof of ownership over their DID. IOTA Identity also supports selective disclosure mechanisms following IETF standards, including both SD-JWT and Zero-Knowledge Selective Disclosure (ZKSD).

\major{One of the standout features of the IOTA identity framework is its integration with the IOTA ledger}, a Direct Acyclic Graph (DAG), \major{which} provides unique \major{benefits}:

\begin{itemize}
    \item \textit{Feeless}: \major{Unlike traditional blockchains,} IOTA \major{has no} miners or validators. Messages, \major{including DID Documents, can be} stored without incurring \major{transaction fees. This feeless nature allows the deployment of} SSI applications directly on the main network \major{at no cost, promoting global accessibility and inclusivity}. 
    \item \textit{Ease-of-use}: IOTA Identity \major{is accessible} without any cryptocurrency token. \major{It also provides} simple APIs that allow standardized and flexible \major{functionalities}. \major{Additionally, the framework includes a stronghold to securely manage secrets.}
    \item \textit{General Purpose DLT}: IOTA is a general-purpose DLT, \minor{enabling} SSI \minor{integration} with other features such as payments, data streams, smart contracts, and access control.
\end{itemize}

\subsection{Hyperledger Aries}
Hyperledger Aries \cite{aries} is an open-source project \major{within} Hyperledger \major{ecosystem} that offers a complete \major{suite of} tools and libraries for building decentralized identity applications. It \major{supports} DIDs and VCs, ensuring their secure \major{storage} and presentation \minor{while maximizing} privacy preservation. \major{Aries is designed to be highly flexible, supporting} multiple protocols, \minor{various} credential types, ledgers, and registries, \major{as well as \minor{providing} interoperability tools across different identity systems}.

Developers can create novel applications by \major{integrating} application-specific code that controls the Aries agent. Currently, there are several Aries general-purpose agents, \minor{with} others under active development:

\begin{itemize}
    \item \textit{Aries Cloud Agent - Python}: \minor{Suitable} for all non-mobile applications and has production deployments. \minor{It runs alongside a} controller and communicates using an HTTP interface. The controller can be implemented using any language and the agent embeds Indy-SDK. 
    \item \textit{Aries Framework - .NET}: \minor{Designed for} building mobile and server-side agents, \minor{this framework is also used in production environments}. \minor{Similar to the Python-based agent}, the controller can be written in any language, and \minor{the framework can be embedded} as a library. \minor{It also includes} the Indy-SDK. 
    \item \textit{Aries Static Agent - Python}: A configurable agent that does not use persistent storage, \minor{making it lightweight and straightforward for specific use cases.}
\end{itemize}

\major{Despite its potential, the \minor{development resources} available for Aries agents can be somewhat limited. Hyperledger \minor{provides} an online course for developers,  \minor{which is offered} at certain times \minor{throughout} the year.} 

\major{In terms of security and performance, Hyperledger Aries \minor{stands out as} one of the most robust platforms for decentralized identity systems. However, for testing purposes or certification in process-heavy environments, Aries may not be the optimal choice due to its architectural complexity and the steep learning curve required to effectively use and deploy its agents.}

\subsection{Microsoft Entra Wallet Library}
Microsoft has been actively developing the Microsoft Entra Wallet Library \cite{microsoftsdk}, a novel library to manage DIDs and VCs on iOS and Android \major{platforms}. It \minor{enables} \major{mobile} applications to integrate with the Microsoft Entra Verified ID platform, supporting the issuance and presentation of VCs in compliance with various industry standards, including OpenID Connect, Presentation Exchange, and VCs. 

By default, Microsoft Entra Wallet Library uses distinct DIDs for each interaction with relying parties, ensuring privacy \major{protection} by preventing the correlation of user actions. \major{Moreover, the library automates the retrieval of exchanged VCs directly from the original issuer, streamlining the process and maintaining the integrity of the credentials.} The library supports the following requirements, \major{providing flexibility for verifiers and issuers}:

\begin{itemize}
    \item \textit{GroupRequirement}: \major{Allows verifiers or issuers to request multiple requirements simultaneously, aggregating them into a list.}
    
    \item \textit{VerifiedIdRequirement}: Enables a verifier or issuer \major{to request a specific} VerifiedId, \major{which serves as the primary credential for the user.}
    
    \item \textit{SelfAttestedClaimRequirement}: Allows an issuer \major{to request} self-attested claim, which \major{are} simple string \major{values provided by the user}.
    
    \item \textit{PinRequirement}: Enables an issuer to \major{require} a PIN from the user as part of \major{verification or issuance process}.
    
    \item \textit{AccessTokenRequirement}: Allows an issuer to request an Access Token. \major{This} typically involves the use of an external library.
    
    \item \textit{IdTokenRequirement}: Permits an issuer to request an Identity Token (Id Token). \minor{If the token is} not included in the request, \minor{it must be} obtained \major{through} an external library.
\end{itemize}

\major{Despite \minor{its} potential, Microsoft Entra \minor{Wallet Library has} some limitations. To access developer resources, users must register for the Microsoft 365 Developer Program and provide personal information. In terms of usability, the Microsoft Entra Wallet is primarily optimized for mobile devices, limiting its broader use cases. However, it remains a promising solution for IT admins managing access to apps and resources and for app developers seeking to implement \minor{SSO} functionalities using existing user credentials.}

\subsection{Veramo}
Veramo \cite{veramp} is a JavaScript Framework designed \major{to simplify the use of} DIDs and VCs. \major{Its flexibility} and \major{modularity make} it \major{adaptable for a wide range of use cases and} workflows. \major{At the core of Veramo is the} Veramo Agent, \major{which} leverages a plugin-driven architecture \major{to provide scalability and integration with emerging standards in the verifiable data ecosystem}. 

\major{The} Veramo Agent is responsible for creating identifiers, resolving identifiers, credential issuance, credential revocation, credential exchange, and secret application hot sauce. \major{These functionalities are implemented via plugins, which expand the agent’s capabilities \minor{and ensure adaptability}. Additionally, Veramo is multi-platform, running seamlessly across Node.js, browsers, and React Native, making it suitable for a wide array of development environments.} \major{Veramo further enhances usability with a Command Line Interface (CLI), allowing developers to quickly create DIDs and VCs or run a local cloud agent \minor{directly} from \minor{the} terminal. This streamlines both the development and testing processes, making it an attractive tool for decentralized identity solutions.}

\major{Moreover, Veramo supports selective disclosure \minor{through} a dedicated plugin, \minor{enabling} holders to selectively reveal specific claims within a VC, \minor{thus} protecting sensitive information during credential verification. \minor{While} this feature is still in its beta phase, \minor{it demonstrates} Veramo's commitment to offering robust privacy features \minor{aligned} with emerging IETF standards}

\begin{figure*}[t!]
\centering
\includegraphics[width=0.99\textwidth]{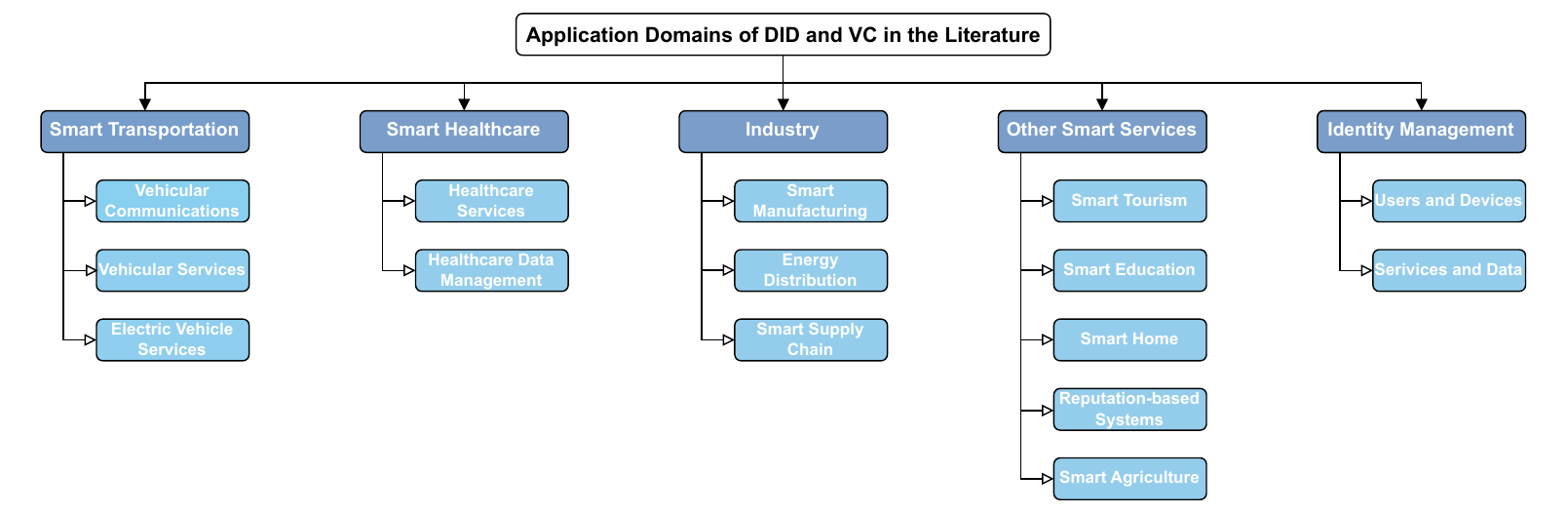}
\caption{Application domains of DID and VC in the existing literature.}
\label{applicationdomains}
\end{figure*}

\subsection{Lessons Learned}
%The exploration of existing implementations and frameworks has highlighted several lessons that shed light on the road ahead. As technology continues to evolve, we anticipate an expanding array of efficient and scalable implementations of DIDs and VCs. Cross-platform implementations, such as DIDKit, offer developers the flexibility to use DIDs and VCs in different environments, including resource-constrained devices and web applications. 

The exploration of the most popular implementations and frameworks highlights their adherence to the W3C standard and open-source nature - \major{two} key \minor{factors favoring} the widespread adoption of DIDs and VCs. In particular, this adherence ensures seamless interoperability across different systems, \major{which is} crucial for the successful integration of these identity technologies. Reviewing implementations shows that some are more generic and web-centric (i.e., DIDKit, Hyperledger Aires, and Veramo), while others are tailored for specific application domains. Consequently, developers must \major{choose} a framework \minor{that} \major{aligns with} their needs and application scenarios. 

Among the existing implementations, DIDKit offers \major{notable} flexibility by supporting a variety of popular programming languages. Its inclusion of a ready-to-deploy HTTP/HTTPS server makes it particularly valuable for developers \minor{approaching} these standards for the first time. Similarly, Veramo is designed to simplify the creation of decentralized identity applications; \major{however,} its limitation to JavaScript may influence developer's decisions. \major{While Hyperledger Aries \minor{provides} a \minor{robust} solution, the initial learning curve may \minor{pose challenges}, especially for developers \minor{unfamiliar} with the Hyperledger ecosystem.}

As outlined throughout this survey, the scope of DIDs and VCs \minor{extends beyond} only individuals \minor{and is} potentially \minor{applicable to any entity, including machines}. In this context, the IOTA Identity framework emerges as a natural and promising solution to \minor{enable the adoption} of these technologies in real-world scenarios, involving \minor{both} humans and \minor{IoT devices.} IOTA's \major{underlying} distributed ledger not only facilitates the use of these technologies but also inherently \major{supports} a VDR, \minor{eliminating} the need for \minor{configuring} \minor{an additional} DLT to disseminate publicly available information like DID Documents.

For developers focusing on mobile-centric solutions, the Microsoft Entra Wallet Library stands out as it is specifically designed for managing DIDs and VCs on various mobile devices (e.g., smartphones and smartwatches) \minor{running} iOS and Android. This implementation empowers developers to provide end-users with comprehensive control over their data through their preferred mobile devices. The active involvement of major technology companies is crucial in advancing these technologies and delivering solutions seamlessly integrated into the modern digital landscape. 
\begin{table*}
    \centering
    \caption{Taxonomy of Smart Transportation Applications.}
\label{tab:transportation_applications}
    \begin{tabular}{>{\centering\arraybackslash}m{2cm}>{\centering\arraybackslash}m{2.5cm} m{0.67\linewidth}c}

\hline
\textbf{Work} & \textbf{Use Case} & \makecell{\centering \textbf{Main Contributions}} \\
\hline
\cite{9303386} & V2X Protocol & A blockchain-based decentralized vehicle-to-anything protocol that enables vehicles to \minor{authenticate and validate} their information. \\
\hline
\minor{\cite{sdp}} & \minor{Secure Data Provenance} & \minor{A protocol that leverages VC to achieve secure data provenance in IoVs.} \\
\hline
\minor{\cite{VDKMS}} & \minor{V2X Protocol} & \minor{A decentralized key management system based on blockchain and SSI principles.} \\
\hline
\cite{9930178} & V2X Reputation System & A V2X reputation system that allows transferring reputation to preserve privacy.\\
\hline
\cite{9747918} & Vehicle Authentication & \minor{A secure registration and authentication mechanism for decentralized VANETs, assisted by double-layer blockchain and DIDs} \\
\hline
\cite{9700631} & Seaport Truck Authentication & A secure, portable, decentralized, and user-controllable identity scheme for truck authentication in seaports. \\
\hline
\cite{9529366} & Vehicle Rights & A decentralized identity management and vehicle rights delegation system.\\
\hline
\cite{9448697} & Vehicle Data & A mechanism for establishing vehicle data provenance in real-time. \\
\hline
\cite{9128642} & Vehicle Identity & A cost-effective implementation of the MOBI standard for vehicle identity.\\
\hline
\cite{KIM2022119445} & Energy Trading & A blockchain-based energy trading scheme for electric vehicle-to-vehicle \minor{interactions}.\\
\hline
\minor{\cite{ssielectric}} & \minor{EV Charging} & \minor{A protocol for EV charging authentication and authorization that can be integrated into ISO 15118.}\\
\hline
\cite{10.1145/3532869} & Customer Privacy & A user-empowered privacy-preserving authentication protocol for \minor{EV} charging.\\
\hline
\end{tabular}
\end{table*}

\section{Applications}\label{application}
In this section, we provide an in-depth analysis of DIDs and VCs, exploring \major{their application} across \major{various} domains. The versatility of these technologies becomes \major{evident through their use} across a wide spectrum of fields, \major{ranging from} smart transportation \major{to} smart healthcare.

\major{We selected the application domains based on the maturity of each field in adopting digital identity solutions, the potential for significant real-world impact using DIDs and VCs, and the existence of a robust body of research supporting these developments.}
Figure \ref{applicationdomains} \major{visually categorizes these application areas, summarizing} key findings from state-of-the-art research that leverage the capabilities of DIDs and VCs. 

\subsection{Smart Transportation}
Recent technological advancements have significantly contributed to the development of modern intelligent transportation systems (ITS) \cite{DIMARTINO2024121949}, which impact various aspects of our lives. Smart transportation involves \minor{integrating} communication technologies \major{with} vehicular services in transportation systems \cite{10136327}. \major{In particular,} Vehicle-to-Vehicle (V2V) communications \minor{have} opened up for novel applications \major{aiming to} improve road traffic efficiency and road safety \cite{jurgen2012v2v}. For example, vehicles can \major{exchange critical data regarding} road conditions or accidents. However, \major{a key challenge in} V2V communications is the lack of trust \major{between} vehicles, \major{as they typically} do not \major{have prior relationships and} rely on centralized network authorities. DIDs and VCs \major{offer promising solutions for establishing} a more secure and decentralized ITS ecosystem. Table \ref{tab:transportation_applications} \major{provides a summary of the relevant literature in this area}.

\begin{figure}[!t]
\centering
\includegraphics[width=0.49\textwidth]{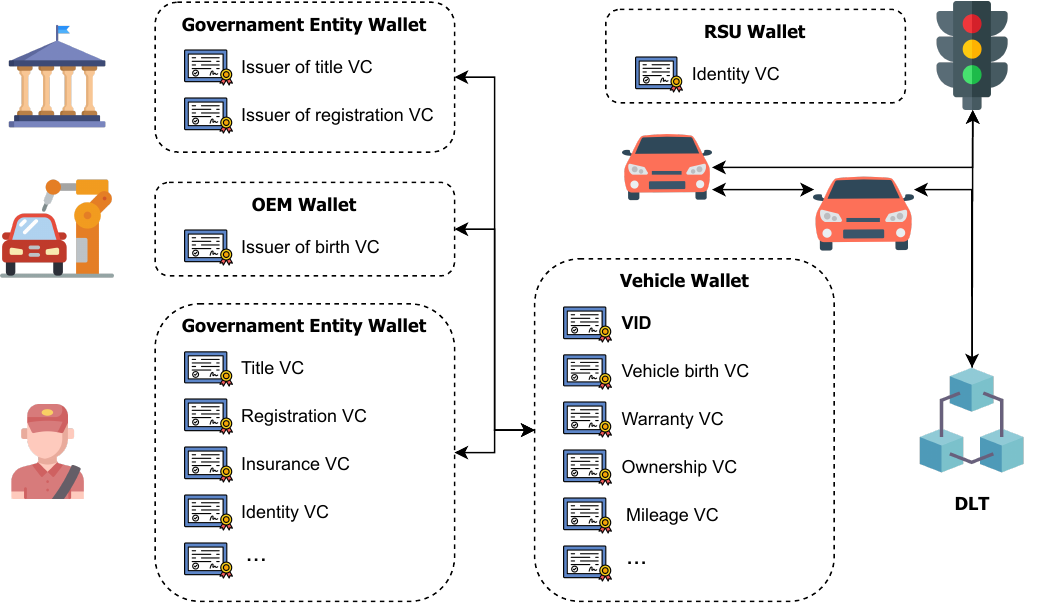}
\caption{The DID/VC-enabled smart transportation reference architecture presented by MOBI [\ref{mobiarch}].}
\label{mobiarch}
\end{figure}

\smallskip
\noindent\textbf{Vehicular Communications.} In V2V communications, DIDs and VCs ensure integrity, authentication, confidentiality, and privacy without \major{relying on a centralized} trusted authority. These technologies, \major{when combined} with DLTs, \major{enable promoting} decentralized solutions that meet the requirements of ITS applications. This trend has been further \major{endorsed} by the MOBI Alliance \cite{mobi}, which introduced a blockchain-based Vehicle IDentification (VID) \major{standard in 2019} \cite{9128642}. \major{Built on the DID specification}, this standard \major{enhances} the \major{traditional} Vehicle Identification Number (VIN), making it compatible with the blockchain ecosystem. Figure \ref{mobiarch} \major{illustrates} a high-level architecture of a potential ecosystem based on VID and DLT. 

\major{An example of such a protocol is} D-V2X \cite{9303386}, a blockchain-based decentralized Vehicle-to-Anything (V2X) protocol that \major{eliminates the need for a trusted intermediary.} The key element \major{of this system} is the decentralized vehicular PKI (D-VPKI), which replaces the \major{traditional} VPKI \major{used} in V2X \major{communications} \cite{vpki}. 
To link a VIN to \major{a} vehicle using \major{the} blockchain, the vehicle \major{first} registers a random DID \major{with} the D-VPKI. The original equipment manufacturer (OEM) \major{then} issues a VC embedding the VIN and any other relevant \major{data}. \major{In a fully decentralized set up,} the vehicle \major{acts as both} the subject and holder. \major{Whenever} the vehicle needs to prove its DID \major{association} with its VIN, it \major{provides} the VC. To preserve privacy, a vehicle registers \major{multiple} DIDs, \major{all} linked to the master DID through ZKP-enabled VCs issued by the OEM, \major{which} will remove \major{the DIDs post-creation}. Alternatively, \major{Secure Multiparty Computation (MPC) can be employed to issue VCs, preventing the OEM from linking the original DID}
\major{Vehicles can also use} pseudonyms \major{by creating} two VPs: one \major{to demonstrate} the link between pseudonym \minor{and} master DID, and \major{another} to \major{validate} the master DID without revealing it. 

In addition, to \major{counter} GPS spoofing attacks, a vehicle \major{can obtain a VC from a nearby} infrastructure (e.g., a traffic light), \major{certifying} its location. \minor{Such a concern can also be addressed through secure data provenance protocols \cite{sdp}. Vehicles register with local Road Side Units (RSU) by presenting a VC attesting \minor{to} some attributes. RSUs maintain a table for each vehicle, \minor{including} its attributes and location at a particular time, leveraging basic safety messages. A vehicle that receives claims from another vehicle, interacts with the RSU to check whether the claimed whether the claim location is correct.}
\minor{Vehicular Decentralized Key Management (VDKMS) \cite{VDKMS} is another proposal that leverages DIDs and VCs to achieve secure communication and efficient key management in V2X networks. Similarly to D-V2X, authorized vehicles are provided with a VC that binds the DID to vehicle information, including its VIN.}

Privacy-preserving computation is \major{a significant} concern in \major{V2X} communications. \major{One method} to avoid tracking \major{vehicles} consists of changing the blockchain address while \major{retaining} reputations \cite{9930178}. This mechanism prevents an external observer from \major{linking new and old addresses. The vehicle locally generates} a new address using a secret, derived from the private key of its DID. \major{These} addresses form a deterministic chain, \major{allowing reconstruction} for auditing or investigation purposes. To transfer reputation, the vehicle \major{generates} a "promise" \major{containing the new} address, the reputation value, and a random nonce. \major{This} information is hashed and sent, along with a ZKP, to a smart contract that verifies the proof, checks the reputation balance, and stores the hash in a Merkle Tree. The old address is \major{then} removed from the reputation mapping, \major{and} the reputation is released to whoever can prove \major{ownership of the new address}.  

%Xiao et al. \cite{15925608520220901} proposed an autonomous and controllable distributed authentication scheme based on DIDs that ensures the credibility of the vehicle's identity. In this scheme, CAs issue VCs, embedding vehicle information, to each registered vehicle. Furthermore, they also generate a secret key pair for each vehicle, including the public key in the DID document stored on the blockchain, and return the DID identifier to the vehicle. The VCs are kept locally providing the vehicle with full control of its information disclosed to RSUs. 

\smallskip
\noindent\textbf{Vehicular Services.}
The \major{integration} of DIDs and VCs paves the way for novel services \major{in} smart transportation, \major{ranging} from optimal route planning and access to trucks \minor{at} seaports to transferring vehicle rights and V2V payments. \major{For instance,} BDRA \cite{9747918} \major{leverages} DIDs and a double-layer blockchain for secure registration and authentication. 
The upper layer comprises all authorized RSUs, while the \major{lower} layer \major{includes} the RSU and the vehicles \major{within their coverage areas}. During registration, each vehicle generates its own DID and submits it to the covering RSU. RSUs \major{collaborate} with each other to create a unique VC, \major{granting} vehicles access to services, such as optimal route planning or \major{real-time road condition updates}. Furthermore, they enable authenticated communications with other vehicles. When a vehicle receives a message, it authenticates the sender and verifies the sender's DID \major{against} the list of users authorized in the \major{lower layer} blockchain\major{, ensuring the sender} reputation \major{meets} a predetermined threshold. \major{The recipient vehicle} evaluates the message \major{content}, \major{providing} feedback to the RSU, which updates \major{its} reputation \minor{on} the blockchain. 

\major{In seaports, trucks can be equipped with SSI credentials, such as "vehicle in service", issued by trusted authorities.} DIDs and VCs can be used to enhance transportation efficiency where thousands of tons of goods are moved \cite{9700631}. \major{When a truck performs actions like picking up containers, it must provide the necessary VCs to validate its authorization.}

\major{DIDs and VCs also streamline vehicle ownership transfers and service access} \cite{9529366}. \major{A vehicle} owner \major{can use pairwise DIDs, private and not shared on the blockchain, to grant temporary access to a service provider for vehicle checks or to a friend borrowing the vehicle.}

\major{As} connected vehicles and mobility applications \major{increasingly} rely on data from diverse sources, ensuring data provenance \major{becomes critical to \minor{preventing} misuse and manipulation}. DIDs can be \major{employed} to secure data provenance in an automotive data processing chain, \major{as demonstrated in} a use case involving an ML \major{for detecting} hazardous driving situations \cite{9448697}. \major{Here,} data provenance 
\major{is} achieved by appending new links - represented by DIDs - \major{within} a chain of signed and linked data versions. Each data point is signed by \major{its} producer and linked to its previous state, \major{enabling traceability throughout the data lifecycle.}

MOBI is \major{actively} developing a vehicle identity standard to \major{support} blockchain-based use cases such as V2V payments and Vehicle-to-Infrastructure (V2I) payments, automotive insurance, and financing. \major{One practical implementation is the} Connected Vehicle Information Network (CVIN) \cite{9128642}, \major{where} each vehicle is identified through a CVID ID, \major{a DID-based identifier.}

\smallskip
\noindent\textbf{Electric \major{Vehicle} Services.}
In recent years, Electric Vehicle (EVs) \major{have experienced significant growth, driven largely by rising concerns about} climate change and sustainability. However, \major{this} gradual expansion has not been followed by a \major{corresponding increase in} charging \major{infrastructure} \cite{en14164933}. The \major{development} of the smart grid and the Internet of Vehicles (IoVs) \major{has} enabled \major{innovative charging methods such as} Vehicle-to-Grid (V2G) and V2V \major{charging}. 

DIDs, VCs, and DLTs lay the foundation for novel energy trading schemes \major{in} V2V environments \cite{KIM2022119445}. \major{In these systems,} DIDs are used to submit bids and reserve energy, the seller DID is \major{publicly accessible to verify} the legitimacy of the sellers and \major{transactions}. The only information stored on the blockchain is the DID Document. Transaction details, such as cost and the amount of energy exchanged, are securely recorded through VCs. \minor{DIDs and VCs can be seamlessly integrated into ISO 15118-20 \cite{ISO15118-2}, the standard defining communication protocols for EV charging. These technologies offer a robust alternative to the complex, centralized PKI traditionally required by the standard \cite{ssielectric}.}

Another \major{critical} concern for EV charging is \major{protecting customer privacy. DIDs and VCs facilitate anonymous yet verifiable charging services, ensuring user privacy while enabling service providers to confirm} the legitimacy of users before offering the service. \cite{10.1145/3532869}. Users register using a DID and a pseudo-ID, which is employed to generate multiple pseudo-IDs, \major{which \minor{are} then} used \major{for} subsequent charging \major{sessions}. \major{Both} the user and the charging station \major{authenticate} each other \major{by verifying} the VC of the other party. ZKPs \major{are employed to validate} the VC without \major{revealing any additional information} about the user.

\begin{table*}
    \centering
    \caption{Taxonomy of Smart Healthcare Applications.}
\label{tab:healthcare_applications}
    \begin{tabular}{>{\centering\arraybackslash}m{2cm}>{\centering\arraybackslash}m{2.5cm} m{0.67\linewidth}c}

\hline
\textbf{Work} & \textbf{Use Case} & \makecell{\centering \textbf{Main Contributions}} \\
\hline
\cite{9105054} & Immunity Passport & An application that \minor{verifies} COVID-19 antibody test/vaccination certifications.\\
\hline
\cite{abid2022novidchain} & Immunity Passport & A blockchain-based privacy-preserving platform for \minor{issuing and verifying} COVID-19 test/vaccine certificates.\\
\hline
\cite{9699221} & Immunity Passport & A blockchain-based platform for secure sharing and validation of COVID-19 vaccination certificates stored in IPFS.\\
\hline
\cite{10034646} & Rare Disease & A rare disease identity system that facilitates communications among different specialists, simplifying the resolution of patient identities.\\
\hline
\cite{9895264} & Healthcare Data & A privacy-aware access control system for healthcare data based on blockchain and SSI.\\
\hline
\minor{\cite{pujari}} & \minor{Healthcare Data} & \minor{An access control system that allows users to directly grant access to their EHRs.}\\
\hline
\cite{10.1145/3426474} & Federated Learning & A privacy-preserving decentralized learning framework for healthcare systems that leverages VCs to allow clients to participate. \\
\hline
\cite{9829562} & IoMT Devices Authentication & A DID-based authentication system for smart healthcare \minor{systems}.\\
\hline
\cite{9621153} & Personal Data Trading & A blockchain-based personal data trading system that enables users to sell their sensitive information.\\
\hline
\end{tabular}

\end{table*}

\begin{figure*}[!t]
\centering
\includegraphics[width=0.99\textwidth]{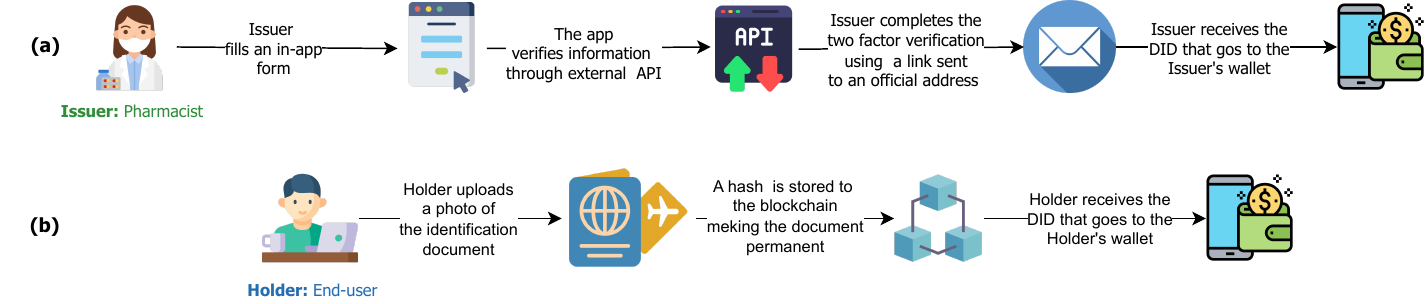}
\caption{Onboarding Process: The pharmacist initiates the onboarding by submitting their information, which undergoes verification through external APIs to generate a DID (a). The user authenticates their identity by uploading a document and receiving a DID (b).}
\label{onboarding}
\end{figure*}

\begin{figure}[!t]
\centering
\includegraphics[width=0.49\textwidth]{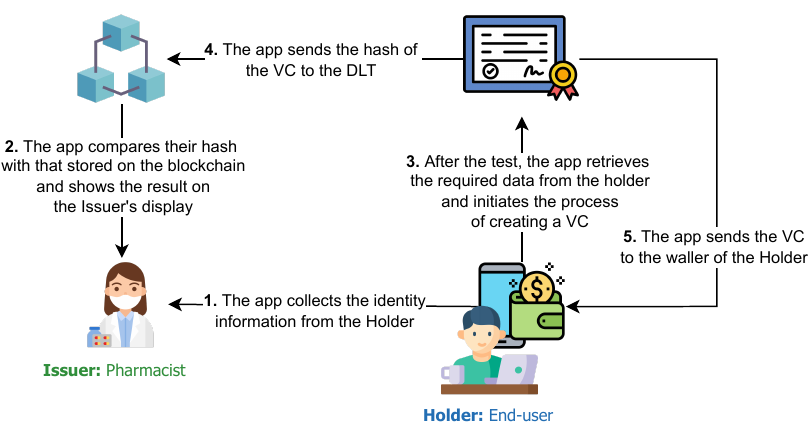}
\caption{Certification Process: The user submits their identity information to the pharmacist, who verifies it. \minor{After} the completion of the medical test, the application issues a VC \minor{containing} the test results.
}
\label{appcert}
\end{figure}

\subsection{Smart Healthcare}
The well-being of individuals has become a central focus in modern society. With the increasing number of healthcare services, the emphasis is on enhancing the health and wellness of patients and elderly individuals, regardless of their location or the time of need \cite{s17020341}. In this context, \minor{adopting} DIDs and VCs offers significant potential \minor{to} address critical security and privacy \major{challenges}. \major{Below}, we summarize \major{key studies from} the literature that cover these aspects. Table \ref{tab:healthcare_applications} reports the contributions of each reference work.

\smallskip
\noindent \textbf{Healthcare Services.}
During the COVID-19 pandemic, the concept of \textit{immunity passport} \major{emerged as a potential means to} slowly come back to normal lives. \major{Among the various proposed technical solutions, those} based on DIDs and VCs \major{have shown the most promise} \cite{Soltani2021}. Mobile apps \major{utilizing these technologies} can provide tamper-proof, privacy-preserving certification for test results and vaccinations \cite{9105054, abid2022novidchain}. In \major{such} systems, \major{trusted entities like} \minor{pharmacies} or national health service, \major{providing VCs for essential documents} such as test results (e.g., blood tests) \major{or} vaccination certificates. These \major{credentials are} securely stored in \major{citizen} hardware or digital wallets \major{after} verification and medical testing \minor{are completed}. \major{Verifiers are responsible for checking the VCs} to determine eligibility for participation in social activities, enforcing necessary \major{health restrictions,} and ensuring that only individuals meeting the required criteria \major{are allowed access}. Figure \ref{onboarding} illustrates the onboarding process for issuers and users, while Figure \ref{appcert} shows the main data flows of certification. 

\major{Some} solutions \cite{9699221} \major{also} \major{incorporate decentralized} technologies \major{like} the Internet Planetary File System (IPFS) for decentralized off-chain document storing, \major{as} storing certificates \major{directly on} the blockchain would be extremely expensive in terms of storage and time. Due to the criticality of such information, IPFS stores the encrypted version of VCs, \major{accessible} only \major{to} authorized entities. The IPFS hash \major{is linked} to the citizen DID which \major{serves as} the user key pair. 

These technologies can \major{also} support patients with rare diseases, \major{requiring} treatments \major{from} multiple care providers and specialists. Effective identity management \major{helps streamline care, avoiding} repetitive steps, enabling scalable and collaborative \major{solutions}. RDIS \cite{10034646} is a rare disease identity system that \major{facilitates} communications \major{between} specialists and \major{resolves} patient identities. It \major{involves} 4 types of participants, each identified through a unique digital identifier (UDID), a DID implementation \major{tied to} a digital profile. Verifiers, \major{such healthcare providers}, \major{manually} check documents and attest to users credentials. Consumers are all the entities, \major{like} \minor{specialists, who} need to access the Unique DID document of a patient. The other participants are patients and their delegates, which may be needed in case patients are minors or suffer from some pathology that limits them. Verifiers issue and sign attestations attached to \major{patient's} UDID document, \major{and} consumers verify \major{these} attestations through a smart contract, \major{which confirms the user presence} in the verifier registry. \major{Every interaction is recorded in an} audit log, \major{containing} only the DID and a timestamp.

\smallskip
\noindent \textbf{\major{Healthcare Data Management.}}
DIDs and VCs lay the foundation of SSI systems, \major{enabling} selective disclosure \major{of user}attributes while \major{maintaining} user privacy. \major{One notable application is} DSMAC \cite{9895264}, a privacy-aware access control system for healthcare data \major{built on} blockchain and SSI \major{principles}. DIDs authenticate access requests, \major{with} access \minor{is} granted or denied according to the roles \major{defined in the user VC under normal conditions.} \major{In} emergency \major{situations}, the system \major{seamlessly} \minor{shifts} to Attribute-based Access Control (ABAC) \major{model, where permissions adapt dynamically} to the contextual attributes included in the VC. \major{In this} model, the patient creates the access control policies and embeds them into the DID Document, which is shared on the blockchain. \minor{Access to Electronic Health Records (EHRs) can also be managed using VCs, which represent the patient's consent to share specific information under predefined conditions \cite{pujari}. Thus, individuals can directly issue a VC that determines what resources can be accessed.}

\major{The increasing volume} of healthcare data \major{can be leveraged by Machine Learning (ML) to support medical professionals through predictive analytics for patient's future health status}. However, \major{strict} regulations prohibit the processing or sharing of individual health information without explicit consent, hindering data exchange \major{across} hospitals and clinics. Federated Learning (FL) \cite{10.1145/3429252} \major{offers a} promising and effective paradigm, \major{enabling multiple healthcare providers to collaboratively train ML models without sharing raw data.} In FL, each participant trains a local ML model and \major{exchanges only} model weights, \major{which are further processed}  (e.g., differential privacy) \major{to prevent inference attacks}. Another critical issue \major{in FL} is the lack of trustworthiness among the participants, \minor{which} can be \major{addressed} by \minor{allowing} \major{a trusted authority, such as a hospital,} \minor{to facilitate collaboration} \cite{10.1145/3426474}.

The \major{rise} of the Internet of Medical Things (IoMT) is \major{also revolutionizing healthcare, with networks of} small, power-efficient, and lightweight wireless sensors, \major{enabling} healthcare providers \major{to} \minor{remotely monitor patients} \cite{gope2015bsn}. 
\major{These sensors collect vital health data and assist in making timely medical decisions. For secure data transmission, devices must} be registered \major{using a} DID \major{and} VC, \major{which then generate authentication tokens for interaction with healthcare providers} and identification numbers (e.g., hospital web server) \cite{9829562}. 

\major{Additionally}, DIDs and VCs \major{facilitate the secure trading of} personal data \cite{9621153}. Users can \major{authenticate their identity} through their DIDs \major{and claim ownership of data using} VCs. If the data owner accepts a buyer's request, \major{they issue a VC that grants the buyer permission to use the purchased data.} ABAC models offer highly flexible and fine-grained access control by regulating authorization based on user\major{-defined} attributes, which may include sensitive \major{health} information.

\begin{table*}[!ht]
    \centering
    \caption{Taxonomy of Industry Applications.}
\label{tab:industry_applications}
    \begin{tabular}{>{\centering\arraybackslash}m{2cm}>{\centering\arraybackslash}m{2.5cm}m{0.67\linewidth}c}

\hline
\textbf{Work} & \textbf{Use Case} & \makecell{\centering \textbf{Main Contributions}} \\
\hline
\cite{9165142} & Anti-counterfeiting System & An anti-counterfeiting system for smartphones. \\
\hline
\cite{9024702} & Firmware/software Updates & A system for \minor{securely} registering 5G IoT devices and updating their firmware.\\
\hline
\cite{9831460} & Device Communications & A framework for connecting unknown devices, updating firmware, and providing monitoring features.\\
\hline
\cite{10143976} & Federated Learning & A framework that enables trustworthy federated learning among unknown clients. \\
\hline
\cite{9889268} & Federated Learning & A framework that leverages blockchain and decentralized identifiers to simplify the use of FL in industry 4.0 scenarios.\\
\hline
\cite{LI2019390} & Smart Grid & A distributed energy system incorporating renewable energy generation and heterogeneous end-users (e.g., residential, commercial, and industrial sectors). \\
\hline
\cite{HARTNETT2021279} & Smart Grid & A framework for coordinating distributed energy resources.\\
\hline
\cite{9333894} & Smart Grid & A DID-based attribute-based access control model for the smart grid.\\
\hline
\cite{T2023107544} & Agriculture Insurance &  A framework that enables trusted agricultural IoT data sharing and provides a decentralized oracle-based access control mechanism for smart contracts in agricultural insurance.\\
\hline
\cite{fi13120301} & Food Supply Chain &  A system that provides full visibility of \minor{processes and} food certifications.\\
\hline
\cite{s23041962} & Supply Chain &  A model to track assets among different blockchain-based supply chain systems. \\
\hline
\cite{xia2023trust} & Software Supply Chain &  A blockchain-empowered architecture for software bill of materials.\\
\hline
\minor{\cite{pcf}} & \minor{Product Carbon Footprint Supply Chain} &  A model that enables trustworthy and confidential sharing of product carbon footprint.\\
\hline
\end{tabular}
\end{table*}

\subsection{Industry} 
In recent years, the convergence of information technology (IT) and operational technology (OT) has given rise to a new era known as smart industries \cite{s22010190}. These industries \major{integrate} traditional manufacturing \major{processes} with cutting-edge technologies like artificial intelligence (AI), data analytics, and IoT, \major{driving and} fostering \major{economic} growth \cite{Li2017}. As the demand for intelligent, \major{data-driven} solutions \minor{rises}, the software has become a vital part of product development for modern industries.

\major{In this evolving landscape,} decentralization has \major{emerged as a key focus due to the need for collaboration and inter-connectivity} among stakeholders. DIDs and VCs \major{have attracted significant} attention as a \major{tool for enabling} secure and interoperable data exchange within industries \cite{ren_survey}. 
Research in this domain \major{has identified four primary areas where DIDs and VCs hold particular promise:} smart manufacturing, \major{energy distribution}, smart agriculture, and smart supply chain. Table \ref{tab:industry_applications} summarizes the main contributions of each referenced \minor{study} \major{in these areas}.

\smallskip
\noindent \textbf{Smart Manufacturing.}
Smart manufacturing \major{leverages advanced technologies to optimize} and \major{automate} industrial processes \cite{VENANZI2023103876}. However, centralized approaches often \major{suffer from limitations in} flexibility, efficiency, and security, \major{while trust issues with} third-party authorities \major{exacerbate these concerns}. To address these challenges \major{the integration of} DIDs and VCs \minor{represents} \major{a compelling solution}. For instance, DIDs and VCs \major{can be used} to combat the counterfeiting of smart devices, such as smartphones \cite{9165142}. \major{In this scenario,}  manufacturers generate a unique DID for each smartphone, \major{using the} International Mobile Equipment Identity (IMEI) number as \minor{an} attribute. Specialized DID methods \major{enable verification of the} IMEI DID document and ownership management, while VCs serve to prove the device ownership.

\major{The \minor{use} of VC is also increasing in the context of Digital Twin (DT) technology for realizing smart manufacturing and Industry 4.0. DTs, characterized by the seamless integration between the cyber and physical spaces, provide a virtual representation of physical objects, processes, or services by harnessing real-time data from their physical counterparts \cite{9882337}. The primary objective of DTs is to conduct temporal and predictive analyses, enabling organizations to derive insights that facilitate informed decision-making and enhance operational efficiency \cite{app11093750}.
In this context, addressing security concerns is crucial, as data integrity in DTs must be guaranteed. Authentic data, trusted sources, and traceable ownership are essential for effective DT implementation. Robust cryptographic measures are required to protect data during transit, thwarting unauthorized tampering. The SIGNED framework \cite{10078428} exemplifies how DLT and VCs can enhance data security in digital twin environments. Each functional unit, \minor{such as a device or software module,} is equipped with a wallet for seamless VC sharing. When issuing a VC, all relevant attributes are encapsulated within a claim, encrypted through a shared secret, and subsequently stored within the VC. This ensures that verifiers can ascertain the integrity of attributes and detect any tampering, while the issuer selectively shares information based on specific requests.}

Furthermore, DIDs and VCs can \minor{facilitate} firmware \major{updates}, as demonstrated by the Gnomon framework \cite{9024702}. IoT devices \major{register} their DIDs with an identity hub before delivery, \major{and receive VCs to manage software updates}. The software publisher issues a new VC to the identity hub, and the device requests the latest software through its VC, \major{verifying} if the software publisher has \major{authorized the update}. If the VC \major{references} a newer version, the device uses the URL included in the VC to download the update. A DLT can \major{further enhance}, firmware updates, and device monitoring \cite{9831460}. Each device is assigned a DID, certified by the associated DID Document \major{that defines its} communication channel. A \major{smart contract-based supervisor manages} logging operations and maintains a list of \major{trusted} devices. Manufacturers can indicate \minor{a} \major{secure} endpoint in the DID Document to enable \minor{trustworthy} firmware downloads.

DIDs and VCs \major{also foster} collaboration among unknown participants \minor{in FL}. \major{For instance,} TruFLaaS \cite{10143976} offers FL as a service, enabling companies to collaborate on \major{shared} tasks, \major{such as predicting machine breakdowns to reduce maintenance costs and enhance production capacity}. Participation is regulated through VCs issued by the service provider, \major{with the option to revoke credentials} \minor{when the quality} of \minor{the} submitted models \minor{drops a certain threshold}. \major{Similarly, FlowChain \cite{9889268} \minor{promotes the adoption of} FL in Industry 4.0 by using DIDs to uniquely identify participants and regulate Industrial Internet of Things (IIoT) device participation in FL training.}

\begin{figure}[!t]
\centering
\includegraphics[width=0.49\textwidth]{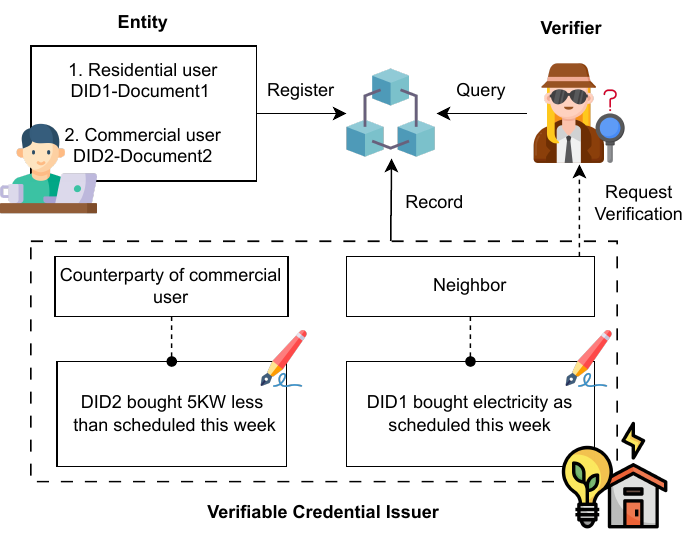}
\caption{VCs can be used to determine the user's credit. In the figure, the commercial user with DID$_2$ has bought less energy than expected, whereas the neighboring user with DID$_1$ has requested the scheduled energy.}
\label{fig:usercredit}
\end{figure}

\smallskip
\noindent \textbf{\major{Energy Distribution.}}
In this subsection, we explore other applications related to \major{energy distribution}. DIDs and VCs can be applied in blockchain-based energy management systems to evaluate user credit \cite{LI2019390}. Each user has a DID that can be used to \major{verify} credit or \minor{assess} the credit of others, \major{all} while preserving privacy, and \major{avoiding \minor{the} disclosure of sensitive information}. VCs are issued by energy system operators to \major{grant users} access to the microgrid and P2P trading system. Users can also issue VCs to \major{confirm that a} counterpart has \major{either} supplied power on demand or bought less than scheduled. \major{These} VCs, \major{which form part of the user’s credit history}, are stored on the blockchain, \major{preventing falsification}. \major{As illustrated in Figure} \ref{fig:usercredit}, the \major{collection} of VCs issued to a user constitutes their credit.

\begin{figure*}[!t]
\centering
\includegraphics[width=0.99\textwidth]{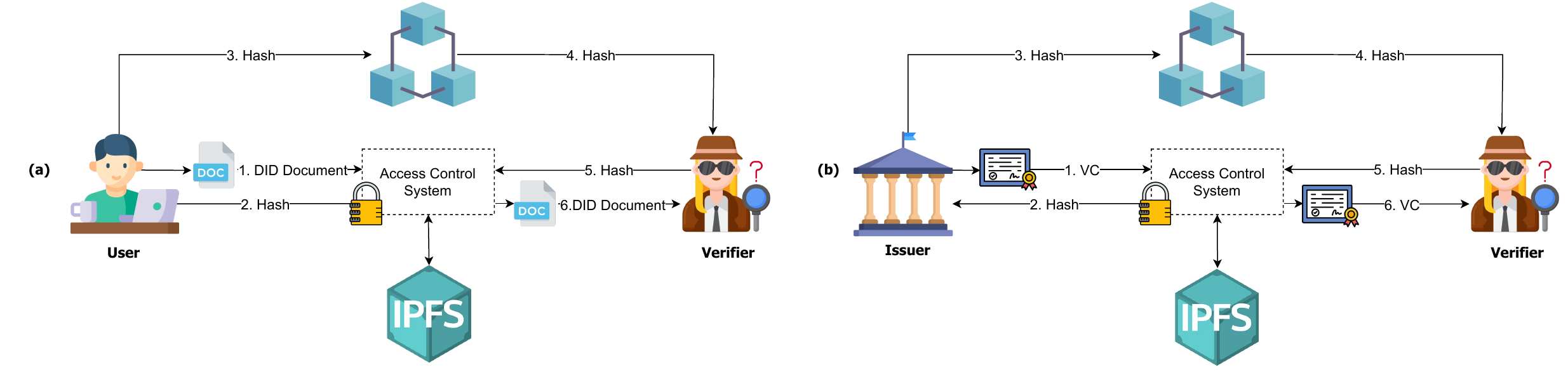}
\caption{General workflow for managing DIDs and VCs using DLT and IPFS. In (a), a user uploads their DID on IPFS and publishes the corresponding identifier (i.e., a hash value) on a DLT; such information is then used by verifiers to retrieve the DID document and access to \minor{the} user's publicly available information. Similarly, in (b), issuers and verifiers perform the same operation referring to VC instead of a DID.}
\label{fig:ipfsdlt}
\end{figure*}

Flex \cite{HARTNETT2021279} \major{provides another example of how DIDs and VCs can enhance energy systems. This framework coordinates distributed energy resources by registering stakeholders through DIDs. These identifiers facilitate interactions between stakeholders}, such as system operators and retailers, \major{while allowing} customers \major{to} present pre-defined documentation to \major{authorized parties for attribute validation}. For example, a hardware vendor can verify the capacity or model number of the inverter of a rooftop solar system. DIDs and VCs allow \major{users to prove their} attributes to any authorized market participant or system without disclosing the underlying information. 

In the \major{context of} smart grids, ABAC models are \major{widely used} for their granularity and flexibility. However, \major{using user} attributes for access control \major{can raise significant} privacy concerns. To address this challenge and mitigate privacy risks, a promising solution is to implement a DID-based ABAC system \cite{9333894}. This approach leverages pairwise DIDs to uniquely identify participants, with each identifier tailored to the specific \major{interactions}. To further enhance privacy protection, traditional user attributes are replaced with VCs, which attest to the validity of certain user attributes without disclosing sensitive information. 

\smallskip
\noindent \textbf{Smart Supply Chain.}
The integration of DIDs, VCs, and DLTs holds significant promise for \major{enhancing} the industrial supply chain within the \major{broader} industry ecosystem. These technologies \major{can significantly improve} transparency, traceability, interoperability, \major{and \minor{logistics} operations}. 

\major{DLTs provide transparency and immutability, which are key to fostering trust} and cooperation between \major{supply chain \minor{participants}, including} companies and manufacturers. \major{With DLTs,} all participants \major{can} access reliable and tamper-proof information, \major{promoting efficient collaboration}. On the other hand, DIDs \major{enable the tracking of} entities involved in supply chain operations, such as IoT devices and \major{personnel}. For instance, each node in the supply chain can require certifications, which are issued as VCs by authorized bodies \cite{fi13120301}. These certifications \major{are} \minor{publicly} accessible and verifiable. When a participant seeks a process certification, an off-chain verification procedure \major{ensures they} meet the \major{necessary} requirements. If successful, the certification body issues a VC stored on the IPFS, \major{with} the resulting hash \minor{being} signed and recorded on the blockchain. \major{Verifiers} retrieve the VC by \major{accessing} the hash from the blockchain and the corresponding data on IPFS. This \major{mechanism} ensures the integrity and authenticity of the certifications, providing a robust trust \major{model} within the supply chain. Figure \ref{fig:ipfsdlt} illustrates \minor{how} \major{this process} \minor{leverages} IPFS and DLTs to manage DIDs and VCs. 

Furthermore, VCs can \major{enhance} interoperability \major{across} different blockchain-based supply chain systems. An example of this is the Verifiable Supply Chain Credential (VSCC) \cite{s23041962}, which extends the traditional VC data model to verify \major{asset alterations} tracked \major{across multiple} blockchains. \major{This} allows seamless \major{data} integration from different systems, \major{enhancing} the overall efficiency and effectiveness of the supply chain. In \major{shipping, for instance, DIDs and VCs streamline} interaction between \major{delivery \minor{agents}} and the mailbox \major{for storing delivered items} \cite{10.1145/3581971.3581992}. The mailbox is identified through an anonymous DID and \major{communicates} its availability to the marketplace. For each purchase, the shipping deliverer is provided by the marketplace with a VC that \minor{allows access to} the mailbox, \major{while the customer receives a VC} as proof of withdrawal. 

Beyond \major{traditional} supply chain \minor{functionalities}, the combination of blockchain and VCs can revolutionize software development, \major{particularly \minor{for} sharing} the Software Bill of Materials (SBOM) \cite{xia2023trust}. Blockchain provides a secure and transparent \major{way} to \major{store and share} data, while VCs ensure the authenticity and integrity of the SBOM information. \major{Oversight authorities \minor{may} issue} VCs certifying that software \minor{vendors adhere} to secure software development standards and practices. These \major{credentials} are then \major{recorded} on the blockchain, enabling stakeholders to easily verify the authenticity of the software and its components, thereby promoting trust in software development processes. 

\minor{VCs can also help achieve trustworthy supply chain exchange for product carbon footprints \cite{pcf}. This is particularly relevant for manufacturers, as selective disclosure enables them to share only specific information from certifications related to the carbon footprint of their products while safeguarding trade secrets—such as supplier details—that are typically included alongside the certified carbon footprint value.}

\begin{table*}
    \centering
    \caption{Taxonomy of Other Smart Applications.}
\label{tab:other_applications}
    \begin{tabular}{>{\centering\arraybackslash}m{2cm}>{\centering\arraybackslash}m{2.5cm} m{0.67\linewidth}c}

\hline
\textbf{Work} & \textbf{Use Case} & \makecell{\centering \textbf{Main Contributions}} \\
\hline
\cite{10.1007/978-981-16-7993-3_42} & Smart Tourism & A smart tourism identity authentication service based on blockchain and DID. \\
\hline
\cite{smarttourism} & Smart Tourism & An analysis of how blockchain might offer novel travel experiences within the tourism domain.\\
\hline
\cite{edrevi} & Smart Education & A systematic literature analysis on the use of digital credentials in higher education institutions.\\
\hline
\cite{9566198} & Smart Education & A mechanism to create secure and machine-verifiable academic credentials. \\
\hline
\cite{9223305} & Smart Education & A decentralized verification solution for higher education certificates.\\
\hline
\cite{bdcc7020079} & Smart Education & Empirical findings from a cross-border pilot on \minor{verifying} education credentials between two universities in Italy and Belgium. \\
\hline
\cite{10031276} & Smart Education & An investigation \minor{of} how digital credentials can be integrated into the SSI ecosystem to overcome challenges of academic networks.\\
\hline
\cite{Mahalle2021} & Smart Home & An OAuth-based authorization and delegation in smart homes for \minor{the} elderly.\\
\hline
\cite{9833873} & Smart Home & A capabilities-based access control system for IoT devices.\\
\hline
\cite{10122540} & E-commerce & A reliable reputation systems for the e-commerce ecosystem.\\
\hline
\cite{10078428} & Digital Twin & A framework based on data ownership and verifiability principles, \minor{aiming to ensure} digital assets in DTs \minor{securely} protected.\\
\hline
\end{tabular}
\end{table*}

\subsection{Other Smart Services}
With the \major{rapid advancement} of cloud and IoT technologies, urban centers \major{are transforming} into smart cities. The smart city concept aims to capitalize on the resources available through ubiquitous devices, offering a wide array of innovative services. A \major{fully} smart city \major{ecosystem encompasses} components such as IoT devices, interconnected networks, robust data storage \major{systems}, and \major{powerful} cloud computing \major{infrastructure to support} service delivery. 

While the potential \major{benefits} of smart cities are \major{clear}, their implementation \major{remains in its early stages, with no city achieving full digitalization yet. However,} by combining DIDs and VCs with other technological breakthroughs, these advancements can \major{significantly enhance} smart city services. Table \ref{tab:other_applications} \major{summarizes the key} contributions of the reviewed works. 

\smallskip
\noindent\textbf{Smart Tourism.}
The tourism sector \major{continuously seeks} innovative solutions that can \major{attract more} visitors \cite{s22041619}. On \major{the} one hand, the development of smart tourism aims to create a
more personalized and ever-adapting experience for tourists while increasing economic, social, and environmental sustainability. On the other hand, \major{the industry has faced challenges such as fraudulent travel agencies and} unqualified part-time guides. 

\major{The integration} of DID, VC, and blockchain \major{technology offers} an effective \major{solution for identity authentication in} smart tourism \cite{10.1007/978-981-16-7993-3_42}. \major{Tourism companies that want} to be authorized \major{can submit} identity requests \major{to regulatory authorities in compliance with national standards. Upon successful verification,} they \major{receive} a DID and VC. \major{To alleviate the burden on the blockchain, IPFS} is used to store the DID Documents and VCs, which certify the legitimacy of the smart tourism organization to offer services. 

Some \major{countries} are reducing their reliance on traditional passports. For example, starting in 2024, Singapore's Airport is \minor{implementing} automated immigration clearance, \major{allowing} travelers to depart without passports \major{using} biometric data \cite{CNN}. Blockchain-based SSI \major{could further enable paperless travel} \cite{smarttourism}. 
A tourist \major{can apply} for a visa by presenting their DID to the consulate, which \minor{releases} \major{it} as a VC. \major{Such VCs allow} the tourist to share identity information when accessing services \major{like} hotel reservations and car rentals that \minor{require} identity \major{verification}.

\smallskip
\noindent\textbf{Smart Education.}
Many \minor{higher} education institutions still rely on paper-based graduation certificates, \major{leading to inefficiencies such as slow, costly, and easily falsified validation processes \cite{edrevi}.} These challenges affect both certificate holders and recruiters. \major{Graduates} may \major{face} difficulties when accessing services such as applying for \major{further studies} or \major{transferring to other} universities, while recruiters may struggle with \major{trust and verifying the authenticity of credential.} \minor{The adoption of DIDs and VCs is a promising solution to} \major{create} digital education passports, offering a secure and trustworthy digital alternative to traditional paper-based certificates.

To achieve trustworthiness, certification systems of academic credentials \major{must} be transparent. In this direction, blockchain is considered the ideal technology. The process of issuing a VC can be modeled as a series of sub-credentials issued over time, with each certification step being recorded on the blockchain. This creates a dependency graph, \major{allowing} verifiers to detect and prevent malicious actions while establishing a publicly verifiable causal relationship between achievements \cite{9566198}. To further enhance trust, \major{the} certification \major{process} should be distributed among different officials, \major{reducing the risk of a small group} compromising the \major{system. Additionally,} blockchain and consortium smart contracts can \major{used} to enforce decentralized verification \cite{9223305}. Higher education institutions can register issued certificates \major{on the blockchain}, allowing recruiters to verify \major{their} authenticity and integrity. \major{In this system,} consortium smart contract maps the DIDs of registered educational institutions to their respective smart contract addresses. When an academic \major{institution joins} the consortium, it must deploy its smart contract and \major{gain approval} from other members. The smart contract deployed by each higher education institute \major{manages} the registration and retrieval of certificate hashes. Recruiters can utilize an application to verify certificates, while higher education institutes can use another application to interact with the smart consortium contract to express their intent to join the consortium.

\begin{figure}[!t]
\centering
\includegraphics[width=0.49\textwidth]{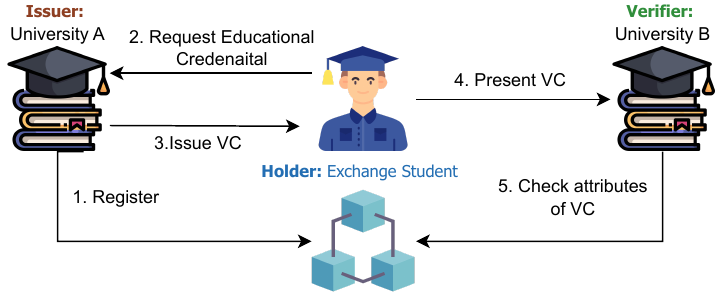}
\caption{A student seeks to participate in an exchange program at University B, supported by a VC issued by University A. To ensure the legitimacy of the credits \minor{earned} at University B, \minor{the} roles \minor{of} issuer (University A) and verifier (University B) \minor{are exchanged}.}
\label{fig:educational}
\end{figure}

\major{Digital credentials also simplify students'} educational services. For example, as shown in Figure \ref{fig:educational}, university admission offices may need to verify a student's prior degrees from other institutions, \major{potentially located} in different countries \cite{bdcc7020079}. \major{Current} academic networks like EMREX use PKI to issue and verify digital credentials. However, linking digital identities to students poses a challenge in EMREX as it is not supported by default. To address this, the ELMO2EDS application \cite{10031276} \major{converts} EMREX digital credentials into EBSI diplomas, a suitable SSI data format that enables user authentication.

\smallskip
\noindent\textbf{Smart Home.}
Smart homes \major{consist of} various interconnected devices that collaborate to offer innovative facilities. These devices communicate through a centralized home gateway, acting as a bridge between the smart home and the Internet. This integration \major{enables} users to remotely control their homes, \minor{offers} \major{increased} convenience and security to their daily lives \cite{CHIFOR2018740}. The home gateway also plays a crucial role in \major{managing user} access to devices \major{by enforcing} predefined access control policies.

\major{In certain situations, external} individuals may require access, such as a technician needing to repair a refrigerator. DIDs and VCs offer a promising solution \major{for secure, temporary access}. Nevertheless, some smart devices may lack the processing capabilities to handle DIDs and VCs directly. To overcome this limitation, the processing of DIDs and VCs can be delegated to an OAuth server \cite{Mahalle2021}. Considering the example of a broken refrigerator, the technician presents their DID and VC, issued and trusted by their company, to the OAuth server. The OAuth server generates a token \major{granting them} that grants access to the refrigerator. In \major{some} cases, smart devices may \major{have the} resources to independently verify VCs, \major{granting access} based on the credential validity. These VCs can include \major{detailed} information about the user capabilities over the device, as \major{demonstrated} in \cite{9833873}.

\smallskip
\noindent\textbf{Reputation-based Systems.}
Nowadays, individuals \major{frequently} share their opinions through \major{online} reviews, expressing preferences for hotels, restaurants, services, and more on platforms like TripAdvisor and Google Reviews. \major{Many} people \minor{also} rely on \major{these reviews} when making \major{decisions}, such as selecting a restaurant or purchasing a \major{product}. However, it is crucial to recognize that these reviews may not always be reliable \major{due to potential} biases, \major{whether} positive or negative. \major{To address this,} VCs can be also leveraged to build more \major{trustworthy} reputation systems. 

For example, \major{in} e-commerce systems,
\cite{10122540} each seller \major{can be} identified through a digital identity and \major{, after a purchase, they can issue feedback tokens as} VCs to customers. \major{These} tokens are used to submit reviews, ensuring the feedback and identity information is authentic and tamper-proof. To \major{encourage} participation, clients \major{are rewarded} with a discount token \major{from the platform after submitting their review.}

\smallskip
\noindent \textbf{Smart Agriculture.}
Agriculture is one of the most valuable sectors of the \major{global} economy. However, recent global warming challenges have \major{significantly impacted food production} \cite{doi:10.1080/10408398.2020.1749555}, increasing the number of people \major{experiencing} acute hunger. The production risks are primarily caused by extreme weather conditions such as heavy storms and \major{prolonged} droughts, \major{which} may cause severe losses in crop yield and lead to marginal farmers’ financial bankruptcy. \major{In response to these challenges,} agriculture insurance \major{has emerged as a} risk management strategy \major{to} support farmers. However, the lack of trustworthiness in agricultural data-sharing platforms and information asymmetries discourage insurers from developing higher-quality insurance schemes. Moreover, smallholder farmers are \major{often} reluctant to insure their crops due to limited awareness, high premium costs, delays \major{in payouts}, and lack of transparency in the \major{claims} process. 

\major{The integration of} DIDs, VCs, \major{and} blockchain can \major{offer} the required guarantees to realize agricultural IoT data sharing \cite{T2023107544}. In this context, smart contracts also play a crucial role, \major{as} they can automate payout to farmers \major{who meet} predefined conditions in agricultural risk assessment. However, \major{due to the deterministic nature of blockchain,} smart contracts cannot \major{directly access data from external sources \minor{such} as} agricultural IoT. \major{This limitation is overcome through the use of oracles} which realizes a bridge between the blockchain ecosystem and the outside world. Each IoT device deployed in farmland is assigned a DID, \major{and} the blockchain \major{serves} as \major{the} source proof \major{for the} VCs issued to \major{these} devices, based on their attributes. 

\begin{table*}
    \centering
    \caption{Taxonomy of Identity and Access Management Applications.}
\label{tab:identity_applications}
    \begin{tabular}{>{\centering\arraybackslash}m{2cm}>{\centering\arraybackslash}m{2.5cm}m{0.67\linewidth}c}

\hline
\textbf{Work} & \textbf{Use Case} & \makecell{\centering \textbf{Main Contributions}} \\
\hline
\cite{9519473} & Identity System & A decentralized digital identity system. \\
\hline
\cite{9387704} & Cross-domain Identity & A decentralized cross-domain identity management system based on blockchain. \\
\hline
\cite{9687671} & Identity System & A distributed identity system for IoT.\\
\hline
\cite{9343033} & Access Control & An access control model for cross-organization identity management. \\
\hline
\minor{\cite{SONG2024267}} & \minor{Selective Disclosure} & \minor{A protocol for the selective disclosure of claims within VCs.} \\
\hline
\cite{10.1145/3538969.3544436} & Identity System & An identity framework that combines SSI and FIDO.\\
\hline
\cite{9472820} & Naming Protocol & A naming protocol for IPFS content identifiers that \minor{enables} self-verifiable content items.\\
\hline
\cite{9472201} & Service Discovery & A P2P discovery system for web services.\\
\hline
\cite{9679363} & Cross-domain Identity & A cross-chain verification model of \minor{DIDs} for JointCloud. \\
\hline
\cite{9481850} & Routing & A protocol enabling routers to self-verify content advertisements.\\
\hline
\end{tabular}
\end{table*}

\subsection{Identity Management}
\major{DIDs and VCs were originally proposed as valuable technologies for identity management. This section reviews systems that leverage these technologies for managing user and device identities, and to identify services and resources.} Table \ref{tab:identity_applications} offers an overview of the referenced papers. 

\smallskip
\noindent \textbf{Users and Devices.}
CanDID \cite{9519473} is a decentralized identity system that enables issuing credentials in a user-friendly manner. The identity subsystem \major{securely} ports identities and credentials from existing web services (e.g., social media platforms and online bank accounts), \major{allowing the creation of} trustworthy credentials without \major{needing to explicitly generate} DID-compatible credentials. CanDID also \major{provides} a key recovery service, \major{enabling} users to retrieve their keys \major{efficiently}.

BIdM \cite{9387704} is a \major{blockchain-based} decentralized identity management system that uses DIDs to identify identities and remove any single-point dependencies. \major{Public keys associated with} each identity \major{are} retrieved \major{through} \minor{the} blockchain-based identity provider, ensuring secure and trustworthy identity management. BIdM \major{employs} a one-way accumulator \major{to} accumulate valid key-value pairs (DIDs and public keys) and \major{support} efficient proof of membership. These accumulator states are stored on a blockchain consortium, \major{preventing} single point of failure and \major{ensuring cross-domain} synchronization.
For cross-domain authentication, the identity-relying party retrieves the authenticated entity information and \major{verifies them} via the accumulator. \major{To enhance user identity representation,} BIdM divides identities into master and shadow: the master identity corresponds to the DID used \major{within} blockchain networks, while the shadow identity is issued by the identity provider and stored offline. SmartDID \cite{9687671} is another blockchain-based distributed identity system \major{tailored} for IoT \major{environments}. It \major{features} a dual-credential mode and a distributed proof system. The dual-credential mode \minor{leveraging} commitments and ZKPs to protect the privacy of sensitive attributes, on-chain identity data, and credential \major{linkages}. 

SSIBAC \cite{9343033} is an access control model based on the XACML specification, \major{integrating} DIDs, VCs, and blockchain technologies to provide privacy and data sovereignty. VCs are encoded in VPs and mapped to access control policies stored in a policy retrieval point. Access requests are regulated through permission validators that bind VPs to attributes, roles, or other data abstractions. \major{Each} VC \major{corresponds to an attribute or role defined} by an access control policy. 

\major{In modern identity and access management systems, privacy preservation is a critical concern. DIDs and VCs offer a novel approach to increase privacy and security by allowing individuals to present credentials without revealing unnecessary data. This is made possible through selective disclosure, a key feature of VCs that enables users to share only a subset of claims while keeping unrelated or sensitive information concealed. For example, when purchasing wine online, a user can disclose only their date of birth to verify their age, while keeping other personal details private. This concept can be extended to other domains, such as healthcare, where a patient might need to prove their insurance coverage without revealing their entire medical history. 
Numerous selective disclosure mechanisms have been developed over time, leveraging advanced cryptographic techniques like ZKPs and selective disclosure signatures to ensure security and privacy protection \cite{surveysd}.} \minor{For example, claims within a VC that represent a diagnostic record can be hidden through HMACs, allowing the patient to selectively disclose data with medical personnel or healthcare service providers \cite{SONG2024267}.}

To \major{further enhance} security, SSI identity frameworks can integrate additional authentication protocols \cite{10.1145/3538969.3544436} like FIDO \cite{srinivas2015universal}. This combination \major{adds an extra} layer of protection, \major{reducing} the risk of \major{data breaches} and preventing unauthorized access, even in the event of identity theft. Each individual uses a FIDO authenticator, \major{which} generates a locally stored key pair serving as their credentials. This key pair is linked to the user DID, facilitating identification and verification of ownership of VCs. \major{For users lacking} external authentication devices \minor{such as} USB tokens, \major{the} Trusted Execution Environment (TEE) \major{can act as the} FIDO authenticator.

\smallskip
\noindent \textbf{Services and Data.} 
IPFS is a decentralized protocol for storing and sharing data, where each item is identified by a content identifier (CID) \major{derived from} a hash function. However, any modification to the data \major{results} in a new CID, which can \major{complicate identifier management} at scale. \major{Integrating} DID and DNS can address this \major{issue} \cite{9472820}. DIDs are used as content names, \major{while} DNSlink \major{binds} a DID to a CID. The content owner publishes a self-verifiable content item on IPFS, which includes the DID document, proof, signed metadata, and the \major{content} itself. The generated DID is associated with the CID through DNSlink, allowing the \major{the item to be located by its DID. Whenever the CID is updated due to content changes,} the corresponding DNS record is updated \minor{as well}.

Beyond data storage, DIDs can be extended to service discovery in peer-to-peer networks by using them to identify threads as global resources \cite{9472201}. DID and VCs \major{enable} identity verification among multiple entities \major{from multiple} organizations. However, they are usually employed to verify identities within a single blockchain. \major{Facilitating} cross-chain DID verification \major{across} different blockchains can \major{promote} collaboration \major{and trust between} entities. For example, in JointCloud - a novel cloud computing paradigm \major{connecting multiple clouds - managing heterogeneous} VCs issued \minor{by} \major{various} organizations \major{is achieved through} a data structure called a verifiable claim, \major{which can be signed by any entity} \cite{9679363}. To \major{verify claims across blockchains}, the verifier executes cross-chain contracts and, if \major{permitted, retrieves} and verifies the claims. To \major{further} enhance \major{trust}, verifiable claims, signers, and verifiers are assigned a credibility value. 

\major{DIDs can also be applied to} self-verifiable content advertisement in named data networking (NDN)  \cite{9481850}, \major{where content is routed based on} identifiers \minor{instead of} \major{network locations. In this model, content owners} use hierarchical DIDs as content names and authorize a publisher to advertise a content name prefix to a specific router. To secure \major{the} advertisement \major{process}, the content owner generates a DID document \major{containing} the JWK representation of a key, \major{which allows} the publisher with that key to advertise \major{the} content.

\subsection{Lessons Learned}
The review of the existing body of research outlines that DIDs and VCs have found applications across seemingly unrelated fields, including smart transportation and smart healthcare. Despite the \minor{differences}, these technologies are used as integral components in realizing SSI systems \minor{across most domains}.

A key feature of DIDs and VCs is \minor{the} capacity to empower individuals with complete control over their data, enabling selective information sharing. Consequently, communication becomes more secure and the risk of potential data breaches is reduced, contributing to building a decentralized and safer digital landscape. Moreover, as underscored by the differences between application domains, these digital identities are utilized across a multitude of platforms and applications, thereby favoring interoperability and improving user experience. Notably, VCs can seamlessly enable access to both educational and healthcare services.

While existing literature primarily focuses on DIDs and VCs as building blocks for SSI,
their ability to provide decentralized verification and the use of DLTs extends beyond, paving the way for fulfilling the objectives of various applications. As highlighted in the literature review, DIDs and VCs can be leveraged in many scenarios where decentralized verification and/or immutability are required, \minor{with} human involvement \minor{not being the primary focus}. For instance, in vehicular scenarios, these technologies are used to enable secure and verifiable communications between vehicles and roadside units \cite{9303386}. Similarly, the Gnomon framework \cite{9024702} employs DIDs and VCs to securely update firmware on devices.

Finally, as discussed in Section \ref{implementation}, there \minor{are connections} between the chosen implementation and the application domain. For example, referring to the frameworks reported in Table \ref{tab:implementations}, in contexts such as smart agriculture and smart transportation, where entities like smart sensors and vehicles require digital identities, IOTA Identity stands as a natural solution. On the other hand, for individuals who prefer managing their digital credentials through mobile applications, e.g., storing medical certificates or travel information, the Microsoft Entra Wallet offers all the necessary functionalities. 
\section{Regulations, Projects \& Organizations}\label{regulation}
Recently, we have been witnessing a growing interest in DIDs and VCs, as demonstrated by the increasing number of regulations, projects, and consortiums emerging worldwide. To provide a comprehensive overview of the global landscape, this section \minor{analyzes} the key initiatives that have emerged since 2019. These represent collaborative efforts among \minor{various} stakeholders, \minor{such as} government entities, technology firms, research institutions, and standardization bodies.

\subsection{Europe}
Europe has always been at the forefront of developing innovative identity management solutions that enable European citizens to use electronic identities (eIDs) for online authentication and communication with online services provided by Member States \cite{app122412679}. This interest in decentralized digital identity \minor{predates} the rise of DIDs and VCs. In 2014, the European Commission recognized the importance of digital identity with the publication of Regulation 910/2014 on electronic identification and trust services (eIDAS) \cite{eidas}. This regulation laid the groundwork for European citizens to securely access online services across the European Union (EU) through a standardized set of digital identity credentials. 

\smallskip
\noindent \textbf{eIDAS 2.0.}
The eIDAS regulation is based on the federated identities model, wherein users are registered with an identity provider. This provider enables them to seamlessly share their identities with service providers that place trust in \minor{that} identity provider. Although this approach provides an SSO experience, it has raised privacy concerns \minor{because} users do not have direct control over their data, and the identity provider has the potential to aggregate information from multiple services, potentially profiling clients.

To address such challenges, the European Commission proposed an updated version of eIDAS in 2021 known as eIDAS 2.0 \cite{eidas2}. This \minor{updated} version shifts from a federated to an SSI model aiming to empower users with direct control over their information, sharing only the minimum amount of data required by the service provider to fulfill \minor{their} request. 

\minor{In May 2024, the EU Regulation 2024/1183 \cite{eu2024regulation1183} introduced the European Digital Identity Framework, designed to strengthen secure and user-controlled digital identities. A key element of this framework is the European Digital Identity Wallet (EUDIW) \cite{9842532}, which is expected to catalyze a major shift toward digital identity, with 500 million smartphone users anticipated to regularly use the EUDIW by 2026 \cite{gartner2024digitalidentity}. This wallet aims to offer European citizens secure access to both public and private services, online and offline, across Europe through an SSI system. It will simplify daily activities for citizens and businesses by supporting various use cases, such as opening a bank account or obtaining educational credentials.}
In this ecosystem, each EUDIW is associated with a DID, and the credentials stored in the EUDIW \minor{take the form of} VCs issued by eIDAS issuers. The verification keys must be publicly accessible by anyone for any use case, ensuring transparency and interoperability.

\smallskip
\noindent \textbf{eSSIF-LaB.}
The European Self-Sovereign Identity Framework Lab (eSSIF-LaB) \cite{essif} is an EU-funded project that focuses on accelerating the adoption of SSI. This \minor{initiative represents a} next-generation digital identity solution, \minor{aiming} to revolutionize electronic transactions both online and \minor{offline. It achieves this} by offering a secure, open, and trustworthy framework. eSSIF-LaB functions as a collaborative ecosystem that brings together various stakeholders, including governments and enterprises, to simplify the implementation and utilization of SSI technology.

\smallskip
\noindent \textbf{EBSI.}
The European Blockchain Services Infrastructure (EBSI) \cite{ebsi} is a collaborative effort between the EU and the European Blockchain Partnership (EBP) to establish a unified platform involving all 27 EU MSs, as well as Norway and Liechtenstein. Its primary objective is to create a robust infrastructure that supports seamless cross-border services. For example, one significant application of EBSI lies in leveraging DIDs and VCs to streamline the management of educational credentials. To illustrate the potential of EBSI in this context, notable institutions like the University of Bologna in Italy and the University of Leuven in Belgium have recently participated in a pilot study focused on the verification of educational credentials \cite{bdcc7020079}. This study \minor{included} verifiable student IDs and transcripts of student records, which can be reliably verified through the EBSI framework.

\smallskip
\noindent \textbf{Regional Initiatives.}
Besides initiatives directly led by European institutions, there are \minor{other} projects at the regional level \cite{regional}. Among European countries, Finland, Spain, \minor{the} Netherlands, and Germany are \minor{actively} working on the development of national SSI frameworks. Moreover, the Spanish Association for Standardization published the first global standard on decentralized identity management based on DLTs in 2021.

\smallskip
\noindent \textbf{\major{GDPR.}} 
\major{Additionally, DIDs and VCs offer a flexible and privacy-preserving approach to digital identity management, enabling robust compliance with key GDPR data rights.}

\begin{itemize} 
\item \major{\textit{Right to Be Informed}: Individuals have the right to be informed about the collection and use of their data. With VCs, data sharing is fully under the control of the user, who decides when and with whom to share their credentials. This ensures that individuals are always aware of where their data is being transmitted and for what specific purpose it is being used.}
\item \major{\textit{Right to Rectification}: In cases where the information within a VC is incorrect or outdated, individuals can request the issuance of the credential with the corrected information, ensuring that their data remains accurate and up-to-date.} 
\item \major{\textit{Right to Be Forgotten}: Users can revoke a VC at any time, rendering it unusable for future verifications and preventing further use of the credential. Additionally, the selective disclosure technique employed by VCs ensures that only the minimum necessary information is shared for any given transaction, significantly reducing the risk of unnecessary data retention by third parties. Finally, VCs may also leverage cryptographic techniques like ZKP, which allows individuals to prove certain properties without revealing underlying personal data, leading to enhanced data minimization.} 
\end{itemize}

\subsection{North America}
The United States (U.S.) has shown a long-standing interest in trusted digital identities \minor{and initiated} significant developments starting as early as 2011 when the National Strategy for Trusted Identity in Cyberspace (NSTIC) was implemented \cite{Li2022}. In 2016, the Department of Homeland Security (DHS) recognized the potential of blockchain-based digital identity technology, \minor{awarding} grants to enterprises involved in this field \cite{9031542}, including one that funded the establishment of the DID working group within the W3C. Within a year, the DHS began providing financial support, accumulating a total of \$4 million in funding targeted \minor{specifically} at small and medium-sized enterprises in the DID domain \cite{li2022establishing}. 

Over the years, the U.S. government's interest in DIDs and VCs has remained consistent. In June 2023, the DHS further showcased its commitment to advancing this field by publishing a solicitation seeking cutting-edge solutions for a privacy-preserving digital credentialing ecosystem \cite{dhs}. The DHS aims to incorporate these solutions into various components and offices, including U.S. Citizenship and Immigration Services (USCIS), U.S. Customs and Border Protection (CBP), and the DHS Privacy Office (PRIV). The global success and adoption of DIDs and VCs \minor{have fueled} this interest. Specifically, the DHS is looking for privacy-preserving components that can seamlessly integrate with existing credentialing systems utilized by \minor{the department}. Among the valuable technical topic areas \minor{identified by the DHS are} digital wallets and software-based verifier implementations for mobile devices.

The Government of Canada, through its Digital Identity and Authentication Council of Canada (DIACC), has been actively exploring the use of DIDs and VCs. DIACC is a nonprofit coalition of public and private sector leaders \minor{working} together to develop a digital identity framework for Canada \cite{diacc}. One of the key initiatives in this space is the Pan-Canadian Trust Framework (PCTF), which comprises a set of rules, standards, and best practices that define how digital identity and VCs can be used across Canada. \minor{The framework} aims to establish a trusted and interoperable digital identity ecosystem that \minor{facilitates} secure and privacy-enhancing transactions. Furthermore, Canada is actively involved in international collaborations and standards development efforts related to DIDs and VCs. For example, the DIACC participates in the W3C Credentials Community Group. In 2018, Canadian public authorities created the Verifiable Organizations Network (VON) to enable governments and organizations to exchange data with VCs \minor{used} to issue digital licenses, permits, and registration \minor{document} to legal entities \cite{sedlmeir2021digital}.

\subsection{South America}
In South America, Argentina is the most active country working on digital identity projects that incorporate DIDs and VCs. The main digital identity project is DIDI \cite{DIDI}, which aims to improve levels of trust and break down some of the socioeconomic and financial barriers that impede access to quality goods and services for vulnerable populations. For example, in 2021, to \minor{promote} financial inclusion, \minor{the project began} providing farmers in the Gran Cacho region with VCs \minor{documenting} their sustainable practices. Such credentials \minor{contribute to} “climate risk scores” that can be presented to financial institutions looking to promote access to financial credit for rural producers and communities. Recently, the province of Misiones approved a law \minor{allowing} the use of blockchain in government management operations. Therefore, \minor{the region} launched a project to improve the adoption of digital wallets. DIDs and VCs can also facilitate secure and trusted cross-border trade.

South American countries are exploring the use of these technologies to streamline customs processes and enhance trade security. For example, Brazil, Colombia, and Costa Rica are among the main countries involved in Farmer Connect \cite{farmer}, a project that leverages SSI technologies to enable end-to-end traceability across the food and agriculture supply chain while addressing regulatory compliance needs. 

\subsection{Asia}
Although digital identities are very widespread in Asia, many of the existing solutions do not specifically employ DIDs and VCs. Indeed, most of the approaches are still centralized. However, it is worth noting that they could be easily integrated with these technologies, whose potential has been recently emphasized by the Founder CTO of Aadhaar \cite{linkedin}, India's biometric ID system.

South Korea stands out as one of the Asian countries at the forefront of adopting DIDs and VCs with a thriving digital identity ecosystem and numerous ongoing projects \cite{kim2022sampling}. The Korea Financial Telecommunications \& Clearings Institute (KFTC) has introduced a blockchain-based digital ID for financial services, while actively participating in the DID Alliance Korea. In 2020, the city of Busan in South Korea launched the Busan Blockchain ID App, enabling citizens to utilize DIDs for accessing various facilities, such as the "multi-child family love" card that provides benefits to households with three or more children. South Korea has also been actively involved in international standardization efforts related to DIDs and VCs, with the Korea Internet \& Security Agency (KISA) being a member of the W3C Credentials Community Group. 

In late 2023, the Chinese Ministry of Public Security collaborated with the Blockchain-based Service Network (BSN) to introduce RealDID \cite{realdid}, an initiative designed to verify real-name digital identities, encrypt personal data, and \minor{provide} certification. %The RealDID system is expected to issue 5 million IDs \minor{by} 2024.
Furthermore, the Chinese WeBank launched WeIdentity \cite{weid}, a project \minor{aiming} to establish a decentralized identity ecosystem using DIDs and VCs. In Hong Kong, using blockchain, DIDs, and VCs have facilitated the development of innovative platforms like ARTRACX Curator \cite{art}, which establishes digital identities for fine art and collectibles, ensuring proper intellectual property protection and authentication. These initiatives showcase the diverse applications and forward-thinking approaches to digital identity in the region.

Since 2023, Taiwan's government has been actively working with W3C and IOTA to implement decentralized identity solutions. Taiwan DID \cite{twdid} is a service that provides Taiwanese citizens with VCs once their residency has been verified. Taiwan DID can be used for digital content subscription services that offer different pricing and content in multiple countries.

\subsection{Africa}
DIDs and VCs \minor{hold} significant potential in addressing identity management challenges in less technologically advanced countries such as those in Africa. Indeed, many of the identity issues experienced in technologically advanced countries, such as Europe and the U.S., have also recently emerged in the second-largest continent. One major concern is the ease of forging digital certificates, \minor{as} credentials can be easily manipulated using tools like Photoshop. This creates a pressing need to enable easy online verification of credentials for trainees, recruiters, and employers. VCs offer a promising solution in this regard. For instance, Gravity Training, a prominent provider of commercial work-at-height solutions in South Africa, has partnered with Dock Network to embrace the use of VCs \cite{gravity}. Additionally, local initiatives like Diwala have facilitated the use of VCs in more than 50 institutions across \minor{nine} African countries \cite{diwala}. 

Another critical issue is \minor{broadening} and \minor{facilitating} access to financial services. Kiva is a nonprofit organization that aims to enhance financial access for underserved communities. In 2019, Sierra Leone, a West African nation of about
7 million \minor{people}, launched the National Digital Identity Platform (NDIP) that \minor{employs} the Kiva Protocol. This platform leverages DIDs and VCs to enable fast, affordable, and secure identity verification for citizens \cite{kiva}. SSI technology can also be valuable in facilitating birth registration. In Kenya, the use of DIDs and VCs \minor{empower} relatives to interact with health workers through smartphones, enabling efficient birth registration and linking mothers to their babies \cite{freytsis2021development}.

\subsection{Australia and Oceania}
Australia has many ongoing and proposed projects focused on storing VCs issued by the government in device-native or state-government wallets and applications. These initiatives aim to facilitate seamless and secure digital identity verification processes. Additionally, Australian jurisdictions are utilizing existing applications and digital wallets for various use cases with particular emphasis on employment and education. For instance, when a recent graduate applies for a job or a higher degree, they may be required to present VCs as proof of their education \cite{9881839}. However, cross-jurisdictional use cases are \minor{also} emerging, where citizens may be asked to provide credentials issued by one or more state-level jurisdictions.

In New Zealand, the Digital Identity Services Trust Framework (DISTF) Bill was approved at the beginning of 2023, establishing the legal foundation for an open accreditation scheme. This framework \minor{promotes} the widespread use of VCs and digital identity in \minor{everyday life} \cite{dinz}. For instance, the Credentials Verification Service for the Nursing Council of New Zealand (CVS-NCNZ) enables nurses educated and licensed outside the country to have their credentials verified and authenticated \cite{ncnz}. This service plays a crucial role in allowing qualified nurses to work \minor{legally} in New Zealand. 

\begin{table*}[!ht]
    \centering
    \caption{Worldwide Initiatives on DIDs and VCs.}
\label{tab:initiatives}
    \begin{tabular}{>{\centering\arraybackslash}m{2cm}>{\centering\arraybackslash}m{2.5cm}m{0.67\linewidth}c}

\hline
\textbf{Government} & \textbf{Initiative} & \makecell{\centering \textbf{Description}} \\
\hline
Europe & eIDAS 2.0 & Enables European citizens to access online services across the EU through SSI.\\
\hline
Europe & EUDIW & Secure storage for European citizens to store credentials and \minor{identity} attributes, and provide them to service providers. \\
\hline
Europe & eSSIF-Lab & \minor{An} EU project \minor{promoting} the adoption of SSI \minor{through} a collaborative ecosystem that \minor{
simplifies its implementation and utilization for governments and enterprises}. \\
\hline
Europe & EBSI & \minor{A} unified platform, involving all 27 EU members Norway, and Liechtenstein, \minor{offering} seamless cross-border services. \\
\hline
Canada & PCTF & A set of rules, standards, and best practices that define how VCs can be used across Canada. \\
\hline
Argentina & DIDI &  \minor{Increases} trust and \minor{reduces} socioeconomic and financial barriers \minor{limiting vulnerable populations' access to quality goods.}\\
\hline
South Korea & Busan Blockchain ID App & Enables citizens of Busan city to use DIDs for accessing smart city facilities. \\
\hline
China & RealDID & A digital identity service for verifying real-name digital identities, encrypting personal data, and certification. \\
\hline
Taiwan & Taiwan DID & Offers digital services based on the residency of users. \\
\hline
Africa & Diwala & Leverages VCs to easily verify credentials. \\
\hline
Africa & NDIP &  Securely identify citizens through DIDs and VCs. \\
\hline
New \minor{Zealand} & DISTF & \minor{A} framework to promote the use of VCs in daily lives. \\
\hline
\end{tabular}

\end{table*}

\subsection{Lessons Learned}
The analysis of emerging regulations, projects, and organizations underscores the global momentum behind DIDs and VCs. These technologies are not only advancing digital identity but also empowering individuals, communities, and organizations across continents. Table \ref{tab:initiatives} \minor{highlights} the main initiatives \minor{worldwide}. 

Europe is undergoing a remarkable digital transformation, moving towards the SSI model, as exemplified by eIDAS 2.0. European citizens \minor{will} be empowered with more control over their data, while the introduction of the EUDIW \minor{ensures the secure} storage and sharing of DIDs and VCs. Moreover, many countries are actively working on national SSI frameworks, recognizing the importance of tailoring digital identity solutions to local needs and regulations. 

In North America, government commitment to promoting the adoption of DIDs and VCs is evident. The U.S. DHS is currently seeking innovative solutions for the privacy-preserving digital credential ecosystem, highlighting the potential of such technologies. The Canadian Digital Identity and Authentication Council is shaping the Pan-Canadian Trust Framework, emphasizing trust and interoperability in the digital identity ecosystem.

Argentina is the most active country in South America. The DIDI project exemplifies how DIDs and VCs can \minor{help} vulnerable populations to \minor{overcome} socioeconomic and financial barriers. The use of VCs to promote climate-responsible agriculture and secure rural credit highlights the real-world impact these technologies can have on local communities.

Among Asian countries, South Korea stands \minor{out} as one of the leaders in adopting DIDs and VCs \minor{for} digital identity solutions. These initiatives reflect the forward-thinking approach to digital identity and its potential to transform various sectors, including finance and public services. In late 2023, China unveiled plans for \minor{introducing} a new digital identity service based on DIDs and VCs, supporting the vision that these technologies will experience a significant boost \minor{in the near future}.

DIDs and VCs are also envisioned as valuable technologies to \minor{foster} inclusivity and \minor{address} the unique challenges that affect Africa. For example, they can enhance financial inclusion and secure birth registration. Local initiatives like Diwala and partnerships with organizations (e.g., Kiva) demonstrate the adaptability and impact of DIDs and VCs in different African contexts. 

Finally, Australia and New Zealand are pursuing the widespread adoption of DIDs and VCs. These initiatives focus on enabling citizens to store and use government-issued VCs, providing secure and seamless identity verification. Cross-jurisdictional use cases demonstrate the potential of VCs to enhance digital identity across borders.

% free trade https://www.developer.tech.gov.sg/our-digital-journey/digital-government-exchange/files/DGX%20DIWG%202022%20Report%20v1.5.pdf

%MOBI south florida health instut university of miami

%https://decentralized-id.com/government/europe/

%\input{sections/lesson}
%\input{sections/summaryfeatures}
\section{Challenges \& Future Research Directions}\label{future}
\minor{Through} the extensive review of \major{DID and VC} applications, we \major{have identified several key} challenges and \minor{outlined} future research directions.  

\subsection{Standardization}
Although DIDs and VCs have been standardized by the W3C, 
there are several challenges \minor{related to} content and protocol standardization across \minor{different domains}. One of the primary concerns is \major{inconsistency in} DID method specifications, \major{which} \minor{vary} in format, completeness, information density, and even versioning \cite{10237038}. These discrepancies \minor{pose significant challenges for} interoperability, as different DID methods may present varying structures, making seamless integration and verification more difficult.

\major{Similar challenges affect the standardization of VCs. While there are domain-specific efforts,} such as the EBSI \major{working on educational credentials for cross-university recognition in Europe \cite{bdcc7020079}, many application domains are still in the early stages of adoption.} 
For example, IoT lacks standardized VC data structures for devices, hindering \minor{interoperability between IoT networks. Moreover, different industries may adopt domain-specific standards, such as VCs for healthcare, leading to fragmentation if they are not designed to be interoperable.}

Standardization concerns \minor{extend to the protocols} \major{that regulate their use}. Many \major{key} aspects, \major{particularly those related to security, remain} non-normative. \major{There is currently no consensus} on cryptographic algorithms, key management practices, or lifecycle management standards for DIDs and VCs. 
\major{Additionally, regulatory and legal frameworks are still evolving, which further complicates interoperability and compliance across different jurisdictions.} \minor{Initiatives like the W3C, ISO standards, and consortiums like Trust over IP (ToIP) \cite{9031548} are working to address these challenges by promoting unified frameworks and conducting interoperability testing to facilitate seamless integration.}

\subsection{\major{Scalability}}
Decentralized architectures based on DIDs and VCs offer greater scalability than centralized systems. This is a natural consequence \minor{because} they do not rely on centralized authorities, which often become \minor{bottlenecks} in traditional systems. DLTs used for storing DID Documents do not \minor{remarkably impact scalability regarding verification, as this occurs off-chain. The DLTs are solely used} to store and retrieve DID Documents \minor{containing issuer and holder public keys}.
Moreover, a verifier \minor{trusting} a subset of issuers can locally store their public keys, \minor{minimizing ledger} interactions. 

However, \minor{with the increasing} adoption of DIDs and \minor{VCs, scalability challenges emerge at the system level, especially in ledger storage and maintenance. The exponential increase in digital identities and their associated DID Documents will place higher demands on the ledger. Although registration and updates occur less frequently than verifications, the expanding ledger may increase maintenance and access costs, leading to potential bottlenecks in query efficiency. Additionally, DLT-based systems can face network congestion during peak activity, slowing down data retrieval and raising operational costs.}

To address \minor{these challenges}, mechanisms \minor{like} indexing, caching, \minor{and sharding} \cite{9146840} can \minor{improve DID Documents retrieval. Indexing can allow faster lookups, while caching can store frequently accessed data locally to minimize interactions with the ledger. Sharding allows splitting the ledger into smaller, more manageable parts, allowing for parallelized access and reducing latency.}

Ongoing research in selective disclosure protocols \cite{surveysd} \minor{aims to minimize storage requirements for holders and the amount of data processed during verification. By 2026, identity owners are expected to manage multiple VCs issued by various organizations \cite{gartner2024digitalidentity}, making it essential to develop mechanisms that enable holders to maintain minimal additional information while selectively disclosing subsets of their claims. These protocols also reduce the data transmitted during verification, alleviating strain on communication channels, minimizing bandwidth usage, and enhancing verification efficiency}

As the volume of issued VCs grows, novel efficient revocation mechanisms \minor{become increasingly important} to ensure that revocation information can be transmitted and stored without overwhelming the network or consuming excessive storage. \minor{Techniques such as cryptographic accumulators and tail files represent promising approaches for optimizing revocation management} \cite{mazzoccausenix}.

\subsection{\major{Usability}}
\major{The complexity of managing new forms of digital identities may pose a significant barrier to their adoption, particularly for average users. 
To address this, by 2026, all EU member states must develop a secure and user-friendly tool that allows European citizens to easily manage their digital identity for accessing public and private services both online and offline (e.g., via Bluetooth or NFC) \cite{10.1007/978-3-031-37586-6_7}. This tool will integrate a wide range of digital documents, from academic credentials to driving licenses \cite{walletusecases}. To promote their broad adoption, digital wallets must offer intuitive interfaces that abstract the complexity of the underlying technology. Individuals should be able to manage their credentials with simple, clear options for sharing data during verifications while retaining full control over their digital identities. Consent management must be user-friendly, with straightforward prompts clearly explaining the implications of sharing personal information.}

To further enhance the user experience, visual workflows and step-by-step guided processes for presenting VCs can make interactions seamless. Additionally, these wallets should support cross-platform interoperability, allowing users to access multiple services without needing different wallets. \minor{Interoperability requires adherence to common standards, such as those established by the W3C for DIDs and VCs. This ensures compatibility between different digital identity systems and avoids fragmentation, enabling both users and service providers to integrate digital identities into their existing workflows efficiently.}

Moreover, automated features for managing the credential lifecycle should be integrated to reduce the manual burden on users. This includes notifications for credential renewals, automatic updates, and alerts for revocation or expiration, streamlining the process for users unfamiliar with digital identity management. Digital identity systems should be designed to operate seamlessly across multiple devices. By synchronizing credentials between mobile, desktop, and other platforms, users can manage and verify their credentials from anywhere, enhancing both convenience and accessibility. \minor{For instance, interoperability between wallets and identity-verification services across platforms and jurisdictions will be critical in creating a cohesive user experience, particularly for travelers or individuals interacting with international services. Wallet interoperability must also account for region-specific legal requirements because differing privacy laws in regions like the EU and the U.S. can influence how VCs are managed and shared, which affects cross-jurisdictional interoperability.}

\subsection{Integration}
\major{DIDs and VCs represent a paradigm shift in digital identity management, \minor{and} their integration with the existing digital identity ecosystem presents both opportunities and challenges. One of the major concerns is the lack of} a collaborative environment where a community of \major{trusted} issuers actively participates. \major{Currently,} only a limited number of entities are ready to issue VCs. \major{Although traditional issuers of digital credentials may initially be unwilling to issue signed variants like VCs \cite{9519473}, the potential of these technologies could foster greater collaboration and engagement from them.} 

\major{DIDs and VCs have the potential to significantly enhance digital identity ecosystems. For example, social media could leverage these technologies to create verified accounts, addressing concerns about bot activity and creating a safer online environment \cite{Yang_Varol_Hui_Menczer_2020}. In this scenario, VCs play a key role by enabling individuals to prove the authenticity of the information required to create an account. Notably, by employing selective disclosure, only the necessary information would be revealed, preserving user privacy. Similar considerations can be extended to other digital identity environments. DIDs and VCs could eventually replace traditional government-issued IDs. Governments could use VCs to provide secure, tamper-proof identity verification while giving citizens greater control over their data.}

\major{However, this transformation also brings technical changes. Identity owners will need a digital wallet to manage their credentials and must undergo a registration procedure that typically involves an initial verification step. Moreover, existing digital identity systems will adapt to support decentralized verification, issuance, and revocation of VCs. For example, service providers would need public keys from a verifiable data registry to directly authenticate the ownership and validity of the presented credentials. While this may face resistance from established identity providers, regulatory frameworks like GDPR could help to overcome such aversion.}

\minor{To ensure seamless integration, interoperability frameworks like DIDComm \cite{10.1145/3658644.3690300} can address challenges arising from differences in various factors, including programming languages, vendors, and networks.
Additionally, developing robust APIs and SDKs will enable wallet providers to integrate seamlessly with various services, enhancing cross-platform compatibility and reducing fragmentation.
} 

\major{In the short term, broad-scale adoption of these technologies can be achieved by ensuring a smooth transition for legacy systems.} Overcoming the inertia associated with legacy \major{infrastructure and enabling seamless interoperability between traditional and decentralized credentialing processes will be key to facilitating the widespread adoption} of DIDs and VCs. \major{Thus, novel approaches that build} VCs from data, available in existing and unmodified services \cite{10.1145/2976749.2978326, 10.1145/3372297.3417239}, \major{will \minor{be} crucial to promote this integration.}

\subsection{Security and Privacy}
\major{As discussed throughout the paper,} although DIDs and VCs offer many security benefits, especially in \major{terms of} privacy and \major{fine-grained control over personal} data, \major{there are some security concerns that} need to be further explored.

\smallskip
\noindent \textbf{\major{DID Management.}} Granting third parties access to manage DID documents introduces potential vulnerabilities that must be \major{carefully addressed to ensure the integrity and security of decentralized identities}. \major{A} risk is the possibility of the DID controller impersonating the DID owner by generating a new key pair and modifying the public key in the DID document. \major{This highlights the need for robust authorization and auditing mechanisms to detect and prevent unauthorized modifications to DID Documents.} 

\major{Currently}, the public key associated with a DID is \major{often} registered to a blockchain, \major{making it accessible} to other nodes of the network. \major{If} the corresponding private key is \major{compromised}, key rotation \major{must be initiated. In such cases,} the DID owner \major{generates a new key pair and publishes the new public key,} signed with the previous one \cite{9334011}. \major{However, the unpredictable nature of key compromises introduces significant security challenges. Regular key pair updates are one way to reduce the vulnerability window, but this presents a fundamental limitation—while it minimizes risk, it does not eliminate it. Key rotation also introduces usability concerns, as frequent updates can be cumbersome for users. Research should focus on flexible key rotation mechanisms that respond dynamically to key compromises, enabling automatic detection and immediate revocation without relying solely on preemptive updates \cite{9583584}. %this paper also outlines security problems
Integrating MFA, such as one-time passwords (OTP) with key rotation, adds an extra layer of security, making it harder for attackers to steal identities even if a private key is compromised.}

\smallskip
\noindent \textbf{\major{VC Revocation.}} Despite the growing interest in VCs, \major{there remains a significant gap in} research focused on developing novel and efficient \major{revocation mechanism}. At the time of this paper, \major{only one proposed specification exists, but it has not yet achieved W3C standard status and} is on the W3C standards track. \major{As described} in Section \ref{sec:threats}, \major{this is the primary method for revoking VCs at present. In addition, there is only one work \cite{mazzoccausenix} that introduces a novel mechanism for efficient revocation of VCs, specifically designed for IoT networks.}

\major{The field of VC revocation presents a range of unexplored research opportunities. The scalability of current approaches, especially in high-throughput environments or those involving constrained devices, like IoT networks, is underexplored. Additionally, novel approaches that leverage cryptographic accumulators and ZKP could further enhance reliability and security across different contexts. Another promising research direction involves revocation solutions interoperable across various decentralized systems.}

\smallskip
\noindent \textbf{\major{Accountability.}} 
Ensuring privacy while simultaneously adhering to existing regulations, such as Know-Your-Customer (KYC) \cite{doi:10.1509/jm.13.0300} and Anti-Money-Laundering (AML) \cite{Shaikh2021}, poses a significant challenge for DIDs. DIDs are designed with privacy preservation as a core objective, but reconciling this feature with the need to screen users presents several complexities. Modern systems leveraging DIDs should meet these regulations and be able to screen users of the system. Specifically, they should be able to identify and validate credentials associated with users. This becomes crucial for screening individuals against various lists, such as sanctions lists, to determine if any preventative actions, like blacklisting, are warranted.

\smallskip
\noindent \textbf{\major{Privacy.}} Research in this domain should explore novel methods that seamlessly integrate privacy-preserving features while \major{ensuring} compliance with regulatory measures. This involves developing cryptographic techniques or privacy-enhancing protocols that \major{enables} DIDs to provide verifiable information without \major{exposing} sensitive details of users. 

\major{A key area for further exploration is selective disclosure within VCs, which allows users to reveal only specific claims without sharing unnecessary information. While SD-JWT represents a widely accepted solution, it still presents some challenges. Notably, the size of the credential grows linearly with the number of claims, which can significantly impact end-user storage requirements. Furthermore, although SD-JWT conceals the content of hidden claims, it discloses the exact number of claims included, which introduces privacy concerns. This detail could be exploited for inference attacks, where adversaries may deduce sensitive information based on the number of undisclosed claims \cite{mehnaz2022your, 8068648}.}

\section{Conclusions}\label{conclusion}
%Digital identification has always been a persistent concern since the inception of the Internet. Over the years, \major{numerous} solutions have been proposed, \major{with a} contemporary shift towards decentralization \major{that prioritizes end-user control and aligns with evolving privacy regulations.} %However, the concept of digital 
%\major{beyond} individuals including \major{entities like} cloud, edge, and IoT resources. 

This article presents a comprehensive survey on DIDs and VCs, two building blocks to authenticate and authorize entities in the modern digital landscape, enabling secure and trustworthy communication. \major{We} analyze DIDs and VCs in terms of \major{threats and mitigation,} implementations, application domains, and regulations. \major{Our analysis} of the available implementations \major{offers} developers valuable insights into the features of existing libraries, \major{aiding informed decision-making across frameworks}. The \major{review} of the application DIDs and VCs demonstrates that their utility goes far beyond SSI systems. Furthermore, \major{we examine} global initiatives and projects \major{led} by governments and political institutions, \major{showcasing the growing adoption of these technologies.} Finally, \major{we discuss emerging challenges and outline future research directions, providing a valuable foundation for ongoing studies in this field.}

\section*{Acknowledgments}
This work was partially supported by the project SERICS (PE00000014) under the MUR National Recovery and Resilience Plan program
funded by the European Union - NextGenerationEU, the U.S.  National Science Foundation's Intergovernmental Personnel Act Independent Research \& Development Program, and Cyber Florida. %, Google, and Microsoft. 
The views expressed by the authors in this paper are their own, not those of the funding entities.
% This work was partially supported via US National Science Foundation's Intergovernmental Personnel Act Independent Research \& Development Program, US National Security Agency (Award: H982302110324), Florida International University Graduate School, Cyber Florida, Google ASPIRE Program, and Microsoft. The views expressed are those of the authors only, not of the funding agencies.

\bibliographystyle{IEEEtran}
\bibliography{main}

%\newpage

\vspace{-33pt}
\begin{IEEEbiography}[{\includegraphics[width=1in,height=1.25in,clip,keepaspectratio]{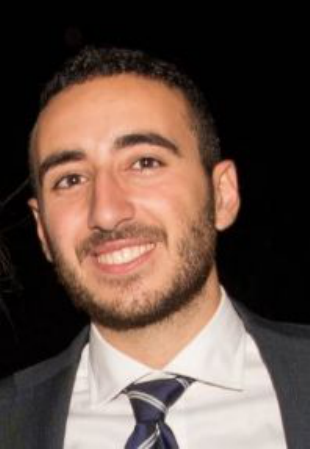}}]{Carlo Mazzocca}
received his Ph.D. in Computer Science and Engineering in 2024 from the University of Bologna, Bologna, Italy. Currently, he is an Assistant Professor in Tenure Track at the University of Salerno, Salerno, Italy. His research primarily focuses on security and privacy aspects, with a particular emphasis on digital identity, security mechanisms built on distributed ledger technologies, authentication and authorization solutions for the cloud-to-thing continuum.
\end{IEEEbiography}

\vspace{-33pt}
\begin{IEEEbiography}[{\includegraphics[width=1in,height=1.25in,clip,keepaspectratio]{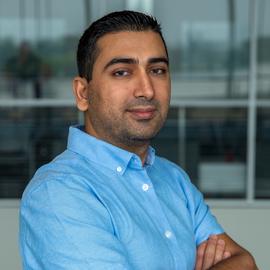}}]{Abbas Acar}
received his MSc and Ph.D. degrees in the Department of Electrical and Computer Engineering at Florida
International University in 2019 and 2020, respectively.
Before that, he received his B.S. degree in Electrical and
Electronics Engineering from Middle East Technical University in 2015. His research interests include continuous
authentication, IoT security/privacy, and homomorphic encryption. More information can be obtained from https:
//web.eng.fiu.edu/aacar/
\end{IEEEbiography}

\vspace{-33pt}
\begin{IEEEbiography}[{\includegraphics[width=1in,height=1.25in,clip,keepaspectratio]{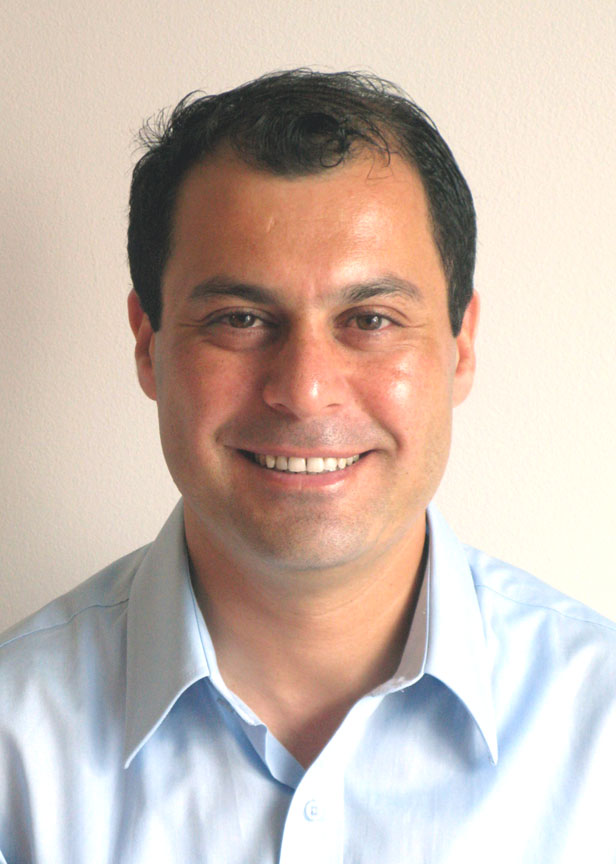}}]{Selcuk Uluagac}
is an Eminent Scholar Chaired
Professor and the director of the Cyber-Physical
Systems Security Lab in the School of Computing
and Information Sciences at Florida International
University (FIU), Miami, Florida, USA. Before FIU,
he was a Senior Research Engineer at Georgia
Institute of Technology and at Symantec. He holds
an M.S. and Ph.D. from Georgia Tech and an M.S.
from Carnegie Mellon University. He is an expert on
security and privacy topics with hundreds of scientific/creative works in practical and applied aspects
of these areas. He received the US NSF CAREER Award (2015), the US Air
Force Office of Sponsored Research’s Summer Faculty Fellowship (2015),
and the University of Padova's (Italy) Summer Faculty Fellowship (2016). His
research in cybersecurity has been funded by numerous government agencies
and industry. He has served on the program committees of top-tier security
conferences such as ACM CCS, IEEE Security \& Privacy (“Oakland”),
NDSS, and Usenix Security, inter alia. He was the General Chair of the
ACM Conference on Security and Privacy in Wireless and Mobile Networks
(ACM WiSec) in 2019. Currently, he serves on the editorial boards of IEEE
Transactions on Information Forensics and Security as Deputy Editor-in-Chief, and IEEE Transactions on Mobile Computing and Elsevier Computer
Networks Journal as associate editor. More information can be obtained from
http://nweb.eng.fiu.edu/selcuk/.
\end{IEEEbiography}

\vspace{-33pt}
\begin{IEEEbiography}[{\includegraphics[width=1in,height=1.25in,clip,keepaspectratio]{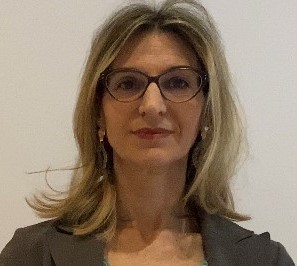}}]{Rebecca Montanari}
is Full Professor at the University of Bologna since 2021 carrying out her research in information security and the design/development of middleware solutions for services in mobile and IoT systems. Her research is currently focused on blockchain technologies to support various supply chains, including agrifood, manufacturing and fashion and on security systems for Industry 4.0. She has established several collaborations with state-of-the-art research centers in the field of semantic web and policy management, such as Nokia Research Center Cambridge, Massachusetts Institute of Technology, Imperial College London, Institute for Human and Machine Cognition, and Pensacola-USA. She has taken part as both coordinator/scientific responsible and participant in several research projects, internationally and in the European Community based in all areas of  ICT, and also funded by Italian Organizations, such as the several Research Ministry and Regional funding systems.
\end{IEEEbiography}

\vspace{-33pt}
\begin{IEEEbiography}[{\includegraphics[width=1in,height=1.25in,clip,keepaspectratio]{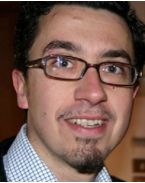}}]{Paolo Bellavista}
received the Ph.D. degree in computer science
engineering from the University of Bologna, Italy,
in 2001. He is currently a Full Professor with
the University of Bologna. His research interests include middleware for mobile computing,
QoS management in the cloud continuum, infrastructures for big data processing in industrial
environments, and performance optimization in
wide-scale and latency-sensitive deployment environments. He serves on the Editorial Boards of IEEE COMMUNICATIONS
SURVEYS AND TUTORIALS, IEEE TRANSACTIONS ON NETWORK AND SERVICE
MANAGEMENT, IEEE TRANSACTIONS ON SERVICES COMPUTING, ACM CSUR, ACM
TIOT, and PMC (Elsevier). He is the Scientific Coordinator of the H2020
IoTwins Project (https: www.iotwins.eu).
\end{IEEEbiography}

\vspace{-33pt}
\begin{IEEEbiography}[{\includegraphics[width=1in,height=1.25in,clip,keepaspectratio]{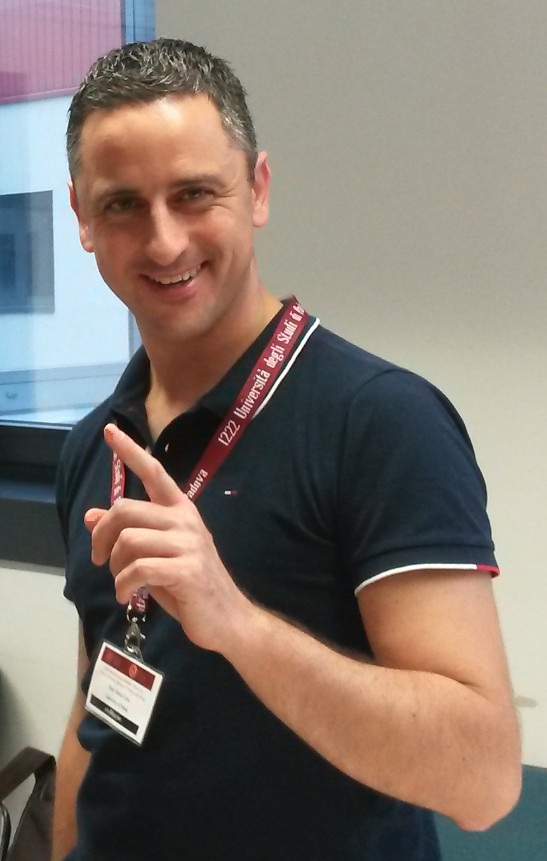}}]{Mauro Conti}
received the Ph.D.
degree from the Sapienza University of Rome, Italy,
in 2009. He is a Full Professor at the University of
Padua, Italy. He is also affiliated with TU Delft and
the University of Washington, Seattle. His research
in the area of Security and Privacy is also funded
by companies, including Cisco, Intel, and Huawei.
He published more than 450 papers in topmost international peer-reviewed journals and conferences.
He is the Editor-in-Chief for IEEE Transactions on
Information Forensics and Security and has been an
Associate Editor for several journals, including IEEE Communications Surveys and Tutorials, IEEE Transactions on Dependable and Secure Computing,
and IEEE Transactions on Network and Service Management. He is a Fellow
of YAE and a Senior Member of ACM.

\end{IEEEbiography}

\vfill

\end{document}